\documentclass{bmcart}

\usepackage{amsthm,amsmath}
\usepackage[numbers]{natbib}
\usepackage[utf8]{inputenc} 

\usepackage[T1]{fontenc}
\usepackage{lmodern}
\usepackage{textcomp}
\usepackage{enumitem}  
\usepackage{xspace}
\usepackage{acro}
\usepackage[autostyle]{csquotes}
\usepackage{microtype}
\usepackage[justification=justified,singlelinecheck=true,font={small}]{caption}
\usepackage{setspace}

\usepackage{amsmath}
\usepackage{mathtools}
\mathtoolsset{showonlyrefs=true,showmanualtags=true}
\usepackage{amssymb}
\usepackage{siunitx}   
\usepackage{gensymb}
\usepackage{units}
\usepackage{amsthm}
\theoremstyle{definition}

\theoremstyle{plain}

\usepackage{aligned-overset}
\usepackage{bm}	
\usepackage{upgreek}

\usepackage{float}
\usepackage{graphicx}  
\usepackage{subcaption}
\usepackage{longtable}
\usepackage{multirow}
\usepackage{longtable}
\usepackage{xltabular}
\newcolumntype{L}[1]{>{\raggedright\arraybackslash}p{#1}}

\usepackage{xcolor}	
\usepackage[section]{placeins}
\usepackage[version=4]{mhchem}
\usepackage{tudacolors}

\usepackage{tikz}  
\usepackage{pgfplots}
\usepgfplotslibrary{external}
{manuscript}
\usepackage{pgfplotstable}
\usepackage{tikz-3dplot}		
\usepackage{ifthen}						
\usepackage{pgffor}
\usepackage{pgfmath}
\usetikzlibrary{datavisualization,calc,positioning,arrows.meta,intersections,calc,bending,decorations.markings,shapes,chains,matrix,scopes,fit,decorations.pathreplacing,spy}
\pgfplotsset{
	every axis plot/.append style={line width = 1.5pt}, 
	every mark/.append style={solid},
	legend style={font=\small,align = left},
	every non boxed x axis/.style={x axis line style={->},xtick align=center},
	every non boxed y axis/.style={y axis line style={->},ytick align=center},
	compat=newest,				
	major grid style = {thick},	
	major tick style = {thick},
	minor grid style = {thin},
	minor tick style = {thin},
	/pgf/number format/precision=4
}
\tikzset{wavelet style/.style={semithick,no marks}}
\tikzset{support style/.style={line width=1.2pt,TUDa-2b}}
\tikzset{support style plot/.style={solid,thick,no marks,TUDa-2b}}
\pgfplotsset{support axis/.style={ytick=\empty,xmajorgrids=true}}
\tikzset{konv style/.style={thick,mark=*}}
\tikzset{normal plot/.style={no marks}}
\pgfplotsset{konv surf style/.style={grid=major,shader=flat,colormap name=viridis}}
\pgfplotsset{konv axis/.style={legend style={at={(1.02,1)},anchor=north west},grid= major,cycle list name=myColorCycleList}}
\usepgfplotslibrary{patchplots}  
\pgfplotscreateplotcyclelist{wavelet color list}{black,TUDa-2b,TUDa-9b}
\pgfplotscreateplotcyclelist{myColorCycleList}{black,TUDa-2b,TUDa-9b,TUDa-3c,TUDa-7b,TUDa-11b}
\pgfplotscreateplotcyclelist{myCycleList}{{black,mark=*},{TUDa-2b,mark=triangle*},{TUDa-9b,mark=diamond*},{TUDa-3c,mark=square*},{TUDa-7b, mark=pentagon*}}
\pgfplotscreateplotcyclelist{myCycleList2}{{black,mark=*},{TUDa-2b,mark=triangle*},{TUDa-9b,mark=diamond*},{TUDa-3c,mark=square*},{TUDa-7b, mark=pentagon*},{black,mark=*},{TUDa-2b,mark=triangle*},{TUDa-9b,mark=diamond*},{TUDa-3c,mark=square*},{TUDa-7b, mark=pentagon*}}

\startlocaldefs

\newif\ifpdf\pdftrue 

\newcommand{\remake}{\tikzset{external/remake next}}

\newcommand{\mathmacro}[2]{\DeclareRobustCommand{#1}{\ifmmode#2\else$#2$\xspace\fi}}

\newcommand{\mathrmbf}[1]{\mathrm{\mathbf{#1}}}

\mathmacro{\R}{\mathbb{R}}
\mathmacro{\Z}{\mathbb{Z}}
\mathmacro{\N}{\mathbb{N}}
\mathmacro{\LR}{L^2(\R)}
\mathmacro{\Lne}{L^2\left(\left[0,1\right]\right)}
\mathmacro{\LOmega}{L^2(\Omega)}

\mathmacro{\Tc}{T_\mathrm{c}}		
\mathmacro{\Bc}{B_\mathrm{c}}		
\mathmacro{\Jc}{J_\mathrm{c}}		

\newcommand{\V}[1]{V_{#1}}
\mathmacro{\Vj}{\V{j}}
\newcommand{\W}[1]{W_{#1}}
\mathmacro{\Wj}{\W{j}}

\newcommand{\VOmega}[1]{V^\Omega_{#1}}

\mathmacro{\phihat}{\hat{\phi}}
\mathmacro{\psihat}{\hat{\psi}}
\mathmacro{\mnull}{m_0}
\mathmacro{\meins}{m_1}
\mathmacro{\Mnull}{M_0}
\mathmacro{\PN}{P_N}
\mathmacro{\Ptil}{\tilde{P}}
\mathmacro{\hn}{h_n}
\mathmacro{\gn}{g_n}
\mathmacro{\ccoef}{c}
\mathmacro{\dcoef}{d}
\mathmacro{\Hop}{\widebar{H}}
\mathmacro{\Gop}{\widebar{G}}
\mathmacro{\cvec}{\mathrmbf{c}}
\mathmacro{\dvec}{\mathrmbf{d}}
\mathmacro{\hvec}{\mathrmbf{h}}
\mathmacro{\gvec}{\mathrmbf{g}}
\mathmacro{\Ntil}{\tilde{N}}
\mathmacro{\Hmat}{\mathrmbf{H}}
\mathmacro{\Gmat}{\mathrmbf{G}}
\mathmacro{\phitil}{\tilde{\phi}}
\mathmacro{\psitil}{\tilde{\psi}}

\newcommand{\phia}[1]{\ensuremath{\phi^a_{#1}}}
\newcommand{\phib}[1]{\ensuremath{\phi^b_{#1}}}

\mathmacro{\util}{\tilde{u}}
\mathmacro{\Lop}{\mathcal{L}}									
\mathmacro{\Bop}{\mathcal{B}}									
\mathmacro{\parOmegab}{\partial\Omega_\mathrm{b}}

\mathmacro{\x}{\mathbf{x}}
\newcommand{\cv}[1][]{c_{\mathrm{V}#1}}
\mathmacro{\Omegat}{\Omega_\mathrm{t}}
\mathmacro{\thetadir}{\theta_\mathrm{dir}}
\mathmacro{\thetaneum}{g_\mathrm{neum}}
\mathmacro{\Gammadir}{\Gamma_\mathrm{dir}}
\mathmacro{\Gammaneum}{\Gamma_\mathrm{neum}}
\mathmacro{\thetatil}{\tilde{\theta}}
\mathmacro{\thetahat}{\hat{\theta}}
\mathmacro{\xtil}{\tilde{x}}
\mathmacro{\anzInner}{\mathcal{J}}
\mathmacro{\anz}{\mathcal{D}}

\mathmacro{\A}{\mathrmbf{A}}
\mathmacro{\M}{\mathrmbf{M}}
\mathmacro{\q}{\mathrmbf{q}}
\mathmacro{\qtil}{\tilde{q}}
\mathmacro{\u}{\mathrmbf{u}}
\mathmacro{\band}{\beta}
\mathmacro{\banda}{\band_\A}
\mathmacro{\bandm}{\band_\M}
\mathmacro{\I}{\mathrmbf{I}}
\mathmacro{\B}{\mathrmbf{B}}
\mathmacro{\b}{\mathrmbf{b}}
\mathmacro{\At}{\A_\mathrm{t}}
\mathmacro{\Mt}{\M_\mathrm{t}}
\mathmacro{\qt}{\q_\mathrm{t}}
\mathmacro{\xivec}{\bm{\upxi}}

\newcommand{\Asf}[1][]{\A^\mathrm{sf}_{\lambda#1}}
\newcommand{\Asfe}[1][]{A^\mathrm{sf}_{\lambda#1}}
\newcommand{\Msfcv}[1][]{\mathrmbf{M}^\mathrm{sf}_{\cv#1}}
\newcommand{\Msfecv}[1][]{M^\mathrm{sf}_{\cv#1}}
\newcommand{\Msflambda}[1][]{\mathrmbf{M}^\mathrm{sf}_{\lambda#1}}
\newcommand{\Msfelambda}[1][]{M^\mathrm{sf}_{\lambda#1}}

\mathmacro{\qsf}{\q^\mathrm{sf}}
\mathmacro{\qsfe}{q^\mathrm{sf}}
\mathmacro{\Bsf}{\B^\mathrm{sf}}
\mathmacro{\bsf}{\b^\mathrm{sf}}

\newcommand{\Afe}[1][]{\A^\mathrm{fe}_{\lambda#1}}
\newcommand{\Afee}[1][]{A^\mathrm{fe}_{\lambda#1}}
\newcommand{\Mfecv}[1][]{\mathrmbf{M}^\mathrm{fe}_{\cv#1}}
\newcommand{\Mfeecv}[1][]{M^\mathrm{fe}_{\cv#1}}
\newcommand{\Mfelambda}[1][]{\mathrmbf{M}^\mathrm{fe}_{\lambda#1}}
\newcommand{\Mfeelambda}[1][]{M^\mathrm{fe}_{\lambda#1}}

\mathmacro{\phiex}{\phi^\mathrm{ex}}

\mathmacro{\L}{L}
\mathmacro{\OmegaS}{\Omega_\mathrm{S}}
\mathmacro{\Sfront}{S_\mathrm{front}}
\mathmacro{\Sback}{S_\mathrm{back}}
\mathmacro{\Sside}{S_\mathrm{side}}
\mathmacro{\Tcal}{\mathcal{T}}										
\mathmacro{\Ncal}{\mathcal{N}}									
\mathmacro{\Ecal}{\mathcal{E}}										
\mathmacro{\Sm}{S_m}
\mathmacro{\Seins}{S_1}
\mathmacro{\bffe}{\nu}
\mathmacro{\bfges}{\kappa}
\mathmacro{\K}{K}

\mathmacro{\anzfe}{\mathcal{I}}

\mathmacro{\Ai}{\mathrmbf{A}^\mathrm{i}}
\mathmacro{\Ab}{\mathrmbf{A}^\mathrm{b}}
\mathmacro{\Aie}{A^\mathrm{i}}
\mathmacro{\Abe}{A^\mathrm{b}}

\mathmacro{\Aqu}{\A^\mathrm{Q3D}_{\lambda}}
\mathmacro{\Mqu}{\M^\mathrm{Q3D}_{\cv}}
\mathmacro{\qqu}{\q^\mathrm{Q3D}}

\mathmacro{\Aaqu}{\A^\mathrm{\acs{aq3d}}_{\lambda}}
\mathmacro{\Maqu}{\M^\mathrm{\acs{aq3d}}_{\cv}}
\mathmacro{\qaqu}{\q^\mathrm{\acs{aq3d}}}

\mathmacro{\Asqu}{\A^\mathrm{SQ3D}_{\lambda}}
\mathmacro{\Msqu}{\M^\mathrm{SQ3D}_{\cv}}
\mathmacro{\qsqu}{\q^\mathrm{SQ3D}}

\mathmacro{\lambxy}{\lambda_{xy}}
\mathmacro{\lambz}{\lambda_{z}}
\mathmacro{\Rmat}{\mathrmbf{R}}
\mathmacro{\cvxy}{\cv[,xy]}
\mathmacro{\cvz}{\cv[,z]}
\mathmacro{\Pmat}{\mathrmbf{P}}
\mathmacro{\s}{\mathrmbf{s}}
\mathmacro{\sfe}{\s^{\mathrm{fe}}}
\mathmacro{\ssf}{\s^{\mathrm{sf}}}
\mathmacro{\sfee}{s^{\mathrm{fe}}}
\mathmacro{\ssfe}{s^{\mathrm{sf}}}

\mathmacro{\ui}{\u_\mathrm{i}}
\mathmacro{\ub}{\u_\mathrm{b}}
\mathmacro{\Aii}{\A_\mathrm{ii}}
\mathmacro{\Aib}{\A_\mathrm{ib}}
\mathmacro{\Abi}{\A_\mathrm{bi}}
\mathmacro{\Abb}{\A_\mathrm{bb}}
\mathmacro{\Mii}{\M_\mathrm{ii}}
\mathmacro{\Mib}{\M_\mathrm{ib}}
\mathmacro{\Mbi}{\M_\mathrm{bi}}
\mathmacro{\Mbb}{\M_\mathrm{bb}}
\mathmacro{\usf}{\u^\mathrm{sf}}

\mathmacro{\hmax}{h_{\mathrm{max}}}
\mathmacro{\qmax}{q_{\mathrm{max}}}
\mathmacro{\zq}{z_{\mathrm{q}}}
\mathmacro{\chiq}{\chi_{\mathrm{q}}}

\mathmacro{\Int}{I}
\newcommand{\VInt}[1]{V^\Int_{#1}}
\newcommand{\WInt}[1]{W^\Int_{#1}}

\mathmacro{\ccoefhat}{\hat{c}}
\mathmacro{\dcoefhat}{\hat{d}}
\mathmacro{\cvechat}{\hat{\mathrmbf{c}}}
\mathmacro{\dvechat}{\hat{\mathrmbf{d}}}
\mathmacro{\Hleftmat}{\Hmat^{\mathrm{left}}}
\mathmacro{\Hrightmat}{\Hmat^{\mathrm{right}}}
\mathmacro{\hleftmat}{\widebar{\Hmat}^{\mathrm{left}}}
\mathmacro{\hrightmat}{\widebar{\Hmat}^{\mathrm{right}}}
\mathmacro{\Hinnermat}{\widebar{\Hmat}}
\mathmacro{\Gleftmat}{\Gmat^{\mathrm{left}}}
\mathmacro{\Grightmat}{\Gmat^{\mathrm{right}}}
\mathmacro{\gleftmat}{\widebar{\Gmat}^{\mathrm{left}}}
\mathmacro{\grightmat}{\widebar{\Gmat}^{\mathrm{right}}}
\mathmacro{\Ginnermat}{\widebar{\Gmat}}

\newcommand{\As}[1][]{\A_{\mathrm{s}#1}}
\newcommand{\Ms}[1][]{\M_{\mathrm{s}#1}}
\newcommand{\qs}[1][]{\q_{\mathrm{s}#1}}
\newcommand{\us}[1][]{\u_{\mathrm{s}#1}}
\newcommand{\Bs}[1][]{\B_{\mathrm{s}#1}}

\mathmacro{\thetahatc}{\thetahat^{\mathrm{c}}}
\mathmacro{\thetahatd}{\thetahat^{\mathrm{d}}}
\mathmacro{\thetahatcvec}{\bm{\hat{\uptheta}}^{\mathrm{c}}}
\mathmacro{\thetahatdvec}{\bm{\hat{\uptheta}}^{\mathrm{d}}}

\mathmacro{\minscale}{j_\mathrm{min}}
\mathmacro{\maxscale}{j_\mathrm{max}}
\mathmacro{\statbasis}{\varXi}
\mathmacro{\dynbasis}{\widehat{\varXi}}
\mathmacro{\phihatvec}{\bm{\hat{\upphi}}}
\mathmacro{\psihatvec}{\bm{\hat{\uppsi}}}
\mathmacro{\D}{\mathrmbf{D}}
\mathmacro{\E}{\mathrmbf{E}}
\mathmacro{\Khat}{\widehat{K}}
\mathmacro{\T}{\mathrmbf{T}}
\mathmacro{\Q}{\mathrmbf{Q}}

\newcommand{\Adyn}[1][]{\A_{\mathrm{d}#1}}
\newcommand{\Mdyn}[1][]{\M_{\mathrm{d}#1}}
\newcommand{\qdyn}[1][]{\q_{\mathrm{d}#1}}
\newcommand{\udyn}[1][]{\u_{\mathrm{d}#1}}
\newcommand{\Bdyn}[1][]{\B_{\mathrm{d}#1}}

\newcommand{\Mdlambda}{\Mdyn[, \lambda]}
\newcommand{\Adlambda}{\Adyn[, \lambda]}
\newcommand{\Mdcv}{\Mdyn[, \cv]}
\newcommand{\Mslambda}{\Ms[, \lambda]}
\newcommand{\Aslambda}{\As[, \lambda]}
\newcommand{\Mscv}{\Ms[, \cv]}

\mathmacro{\tolu}{\tau}

\mathmacro{\fact}{\alpha}
\mathmacro{\numkeep}{\beta}
\mathmacro{\numredef}{\gamma}

\mathmacro{\funsol}{H}

\newcommand{\scalarp}[2]{\left\langle#1\,,#2\right\rangle}		
\newcommand{\innerprod}[2]{\left(#1\,,#2\right)}				
\newcommand{\setprop}[2]{\left\{#1:#2\right\}}					
\newcommand{\abs}[1]{\left|#1\right|}							
\newcommand{\norm}[1]{\left\lVert#1\right\rVert}				
\newcommand{\maxnorm}[1]{\norm{#1}_\mathrm{max}}				
\DeclareMathOperator{\spn}{span}								

\mathmacro{\delt}{\partial_t}
\mathmacro{\delx}{\partial_x}
\mathmacro{\dely}{\partial_y}
\mathmacro{\delz}{\partial_z}
\mathmacro{\delxx}{\partial_{xx}}
\mathmacro{\delyy}{\partial_{yy}}
\mathmacro{\delzz}{\partial_{zz}}
\mathmacro{\nablatwo}{\nabla_{\mathrm{2D}}}
\newcommand{\trans}{{\mkern-1.5mu\mathsf{T}}}						
\mathmacro{\evec}{\mathrmbf{e}}									

\mathmacro{\d}{\,\mathrm{d}}										
\mathmacro{\e}{\mathrm{e}}										
\mathmacro{\i}{\mathrm{i}}										

\newcommand{\ie}{i.e.~}									
\newcommand{\eg}{e.g.~}									
\newcommand{\fig}{Fig.}
\newcommand{\Fig}{Fig.}
\newcommand{\sect}{Sec.}

\newcommand{\qdsf}{\acs{q3d}\textsubscript{sf}\xspace}
\newcommand{\qdpoly}{\acs{q3d}\textsubscript{poly}\xspace}

\pgfkeys{tikz/mymatrixenv/.style={decoration={brace,amplitude=0.5em},every left delimiter/.style={xshift=3pt},every right delimiter/.style={xshift=-3pt}}}
\pgfkeys{tikz/mymatrix/.style={matrix of math nodes,left delimiter=[,right delimiter={]},inner sep=2pt,column sep=1em,row sep=0.5em,nodes={inner sep=0pt}}}
\pgfkeys{tikz/mymatrixbrace/.style={decorate,thick}}

\makeatletter
\let\save@mathaccent\mathaccent
\newcommand*\if@single[3]{%
	\setbox0\hbox{${\mathaccent"0362{#1}}^H$}%
	\setbox2\hbox{${\mathaccent"0362{\kern0pt#1}}^H$}%
	\ifdim\ht0=\ht2 #3\else #2\fi
}
\newcommand*\rel@kern[1]{\kern#1\dimexpr\macc@kerna}
\newcommand*\widebar[1]{\@ifnextchar^{{\wide@bar{#1}{0}}}{\wide@bar{#1}{1}}}
\newcommand*\wide@bar[2]{\if@single{#1}{\wide@bar@{#1}{#2}{1}}{\wide@bar@{#1}{#2}{2}}}
\newcommand*\wide@bar@[3]{%
	\begingroup
	\def\mathaccent##1##2{%
		\let\mathaccent\save@mathaccent
		\if#32 \let\macc@nucleus\first@char \fi
		\setbox\z@\hbox{$\macc@style{\macc@nucleus}_{}$}%
		\setbox\tw@\hbox{$\macc@style{\macc@nucleus}{}_{}$}%
		\dimen@\wd\tw@
		\advance\dimen@-\wd\z@
		\divide\dimen@ 3
		\@tempdima\wd\tw@
		\advance\@tempdima-\scriptspace
		\divide\@tempdima 10
		\advance\dimen@-\@tempdima
		\ifdim\dimen@>\z@ \dimen@0pt\fi
		\rel@kern{0.6}\kern-\dimen@
		\if#31
		\overline{\rel@kern{-0.6}\kern\dimen@\macc@nucleus\rel@kern{0.4}\kern\dimen@}%
		\advance\dimen@0.4\dimexpr\macc@kerna
		\let\final@kern#2%
		\ifdim\dimen@<\z@ \let\final@kern1\fi
		\if\final@kern1 \kern-\dimen@\fi
		\else
		\overline{\rel@kern{-0.6}\kern\dimen@#1}%
		\fi
	}%
	\macc@depth\@ne
	\let\math@bgroup\@empty \let\math@egroup\macc@set@skewchar
	\mathsurround\z@ \frozen@everymath{\mathgroup\macc@group\relax}%
	\macc@set@skewchar\relax
	\let\mathaccentV\macc@nested@a
	\if#31
	\macc@nested@a\relax111{#1}%
	\else
	\def\gobble@till@marker##1\endmarker{}%
	\futurelet\first@char\gobble@till@marker#1\endmarker
	\ifcat\noexpand\first@char A\else
	\def\first@char{}%
	\fi
	\macc@nested@a\relax111{\first@char}%
	\fi
	\endgroup
}
\makeatother

\DeclareAcronym{mra}{short = MRA, long = multiresolution analysis, list=Multiresolution analysis}
\DeclareAcronym{dbn}{short = DB-$N$, long = Daubechies-$N$}
\DeclareAcronym{pde}{short = PDE, long = partial differential equation, list=Partial differential equation}
\DeclareAcronym{ode}{short = ODE, long = ordinary differential equation, list=Ordinary differential equation}
\DeclareAcronym{q3d}{short = Q3D, long = quasi-3-D, list=Quasi-3-D}
\DeclareAcronym{aq3d}{short = AQ3D, long = adaptive Q3D}
\DeclareAcronym{fem}{short = FE method, long = finite element method, list = Finite element method}
\DeclareAcronym{felhc}{short = HE-LHC, long = High-Energy Large Hadron Collider}
\DeclareAcronym{fcc}{short = FCC, long = Future Circular Collider}
\DeclareAcronym{ssm}{short = SSM, long = spectral scaling function method, list=Spectral scaling function method}
\DeclareAcronym{arm}{short = ARM, long = adaptive resolution method, list=Adaptive resolution method}
\DeclareAcronym{1D}{short = 1-D, long = one-dimensional}
\DeclareAcronym{2D}{short = 2-D, long = two-dimensional}
\DeclareAcronym{3D}{short = 3-D, long = three-dimensional}

\endlocaldefs

\begin{document}

\begin{frontmatter}

\begin{fmbox}
\dochead{Research}


\title{Quasi-3-D Spectral Wavelet Method for a Thermal Quench Simulation}


\author[id=jb,
  addressref={aff1},                   
  corref={aff1},                       
  email={jonas.bundschuh@tu-darmstadt.de}   
]{\inits{J.B.}\fnm{Jonas} \snm{Bundschuh}}
\author[id=ld,
  addressref={aff1,aff2},
  email={laura.dangelo@tu-darmstadt.de}
]{\inits{L.A.M.}\fnm{Laura A.M.} \snm{D'Angelo}}
\author[id=hdg,
addressref={aff1,aff2},
email={degersem@temf.tu-darmstadt.de}
]{\inits{H.D.G.}\fnm{Herbert} \snm{De Gersem}}


\address[id=aff1]{
  \orgdiv{Institut für Teilchenbeschleunigung und Elektromagnetische Felder (TEMF)},             
  \orgname{Technische Universität Darmstadt},          
  \postcode{64289},
  \city{Darmstadt},
  \cny{Germany}                                    
}
\address[id=aff2]{%
  \orgdiv{Centre for Computational Engineering},
  \orgname{Technische Universität Darmstadt},
  \postcode{64289},
  \city{Darmstadt},
  \cny{Germany}
}



\end{fmbox}


\begin{abstractbox}

\begin{abstract} 
The finite element method is widely used in simulations of various fields. 
However, when considering domains whose extent differs strongly in different spatial directions a finite element simulation becomes computationally very expensive due to the large number of degrees of freedom. 
An example of such a domain are the cables inside of the magnets of particle accelerators. 
For translationally invariant domains, this work proposes a \acl*{q3d} method. Thereby, a \acs*{2D} finite element method with a nodal basis in the cross-section is combined with a spectral method with a wavelet basis in the longitudinal direction. 
Furthermore, a spectral method with a wavelet basis and an adaptive and time-dependent resolution is presented. All methods are verified. As an example the hot-spot propagation due to a quench in Rutherford cables is simulated successfully.
\end{abstract}




\end{abstractbox}
%

\end{frontmatter}


\section{Introduction}\label{sec:introduct}
A standard \ac{fem} allows to discretize and solve differential equations on arbitrary geometries \cite{braess.2007}. Attaining a sufficient resolution with a standard \ac{3D} FE approach may be intractable within a limited computation time. 
For certain geometries, however, an adapted discretization and simulation method may reduce the computational expense. 
This work focuses on multi-scale problems featuring domains with a high ratio between the extend in one spatial direction, referred to as longitudinal direction, and the extend in the other spatial directions, referred to as cross-section, which are, moreover, the translationally invariant with respect to the longitudinal direction. 
If the cross-section requires a high resolution, because it is \eg very detailed, performing a standard \ac{fem} would be computationally very expensive due to a large number of degrees of freedom. Usually, one would remedy this issue by reducing the model to a \ac{2D} \cite{Bortot_2016aa} or even \ac{1D} \cite{Breschi_2015aa} model by exploiting the symmetry planes. 
Unfortunately, this simplification is no longer possible as soon as the physical behavior breaks the symmetry, \eg for a longitudinally asymmetrical heat source in a thermal heat conduction problem.

The described configuration occurs in superconducting magnets in particle accelerators. These magnets struggle with the quench phenomenon, which is a sudden and local transition from the superconducting to the normal state \cite{Myers.2013}. 
Quenches emerge randomly during continuous operation. 
The affected regions normally cool down again and return to a superconducting state in doing so.
Sometimes, however, a quench leads to a thermal runaway and the magnet can take irreversible damage if protection measures are not taken in time \cite{Myers.2013}. 
Sensitive safety margins lead to a high fault time of the accelerator \cite{Apollonio.2015}. 
The further improvement of the protection systems necessitates the study of the transient effects inside the magnet by numerical simulations \cite{Bortot.2018}. 

An alternative to the FE approach is based on a \ac{q3d} method, that becomes applicable by exploiting the geometrical symmetry of the model \cite{DAngelo.2019,DAngelo.2020}. 
Here, a \ac{2D} \ac{fem} in the cross-section is combined with a \ac{1D} spectral method \cite{Canuto_2006aa} in the longitudinal direction. 
While this approach utilizes FE ansatz functions of lowest order, the order of the spectral element ansatz functions is chosen arbitrarily or appropriately for the problem at hand. 
Hence, the \ac{q3d} method is similar to an anisotropic $hp$-method \cite{Schoeberl_2005aa}. 

In earlier works, harmonic functions \cite{DAngelo.2019} and orthogonal polynomials \cite{DAngelo.2020} have been used as basis functions for the spectral method.
In this work, however, wavelets and the corresponding scaling functions are used as an alternative to these. A comparison of orthogonal polynomials and wavelets is made. 
Wavelets are functions that are defined by properties within the \ac{mra} \cite{Mallat.1989}. 
In this work, the Daubechies wavelets \cite{Daubechies.1992} are utilized.

\begin{figure}
	\centering
	\newcommand{\xcord}{0.704}
	\newcommand{\ycord}{0.575}
	\ifpdf%
	\includegraphics{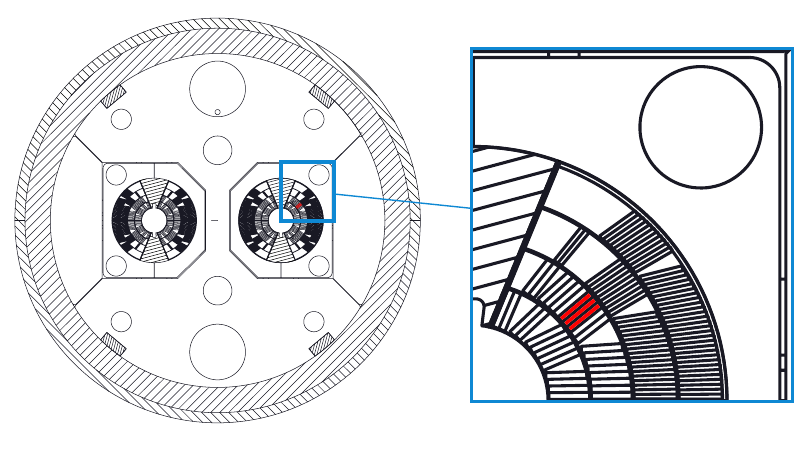}%
	\else%
	\tikzsetnextfilename{magnet}%
	\begin{tikzpicture}[spy using outlines={chamfered rectangle, magnification=6, width=3.26cm, height=3.59cm, connect spies,every spy on node/.append style={very thick}}]
		\node (fig) at (0,0) {\includegraphics[trim = 0 8cm 0 0,clip,scale=.2]{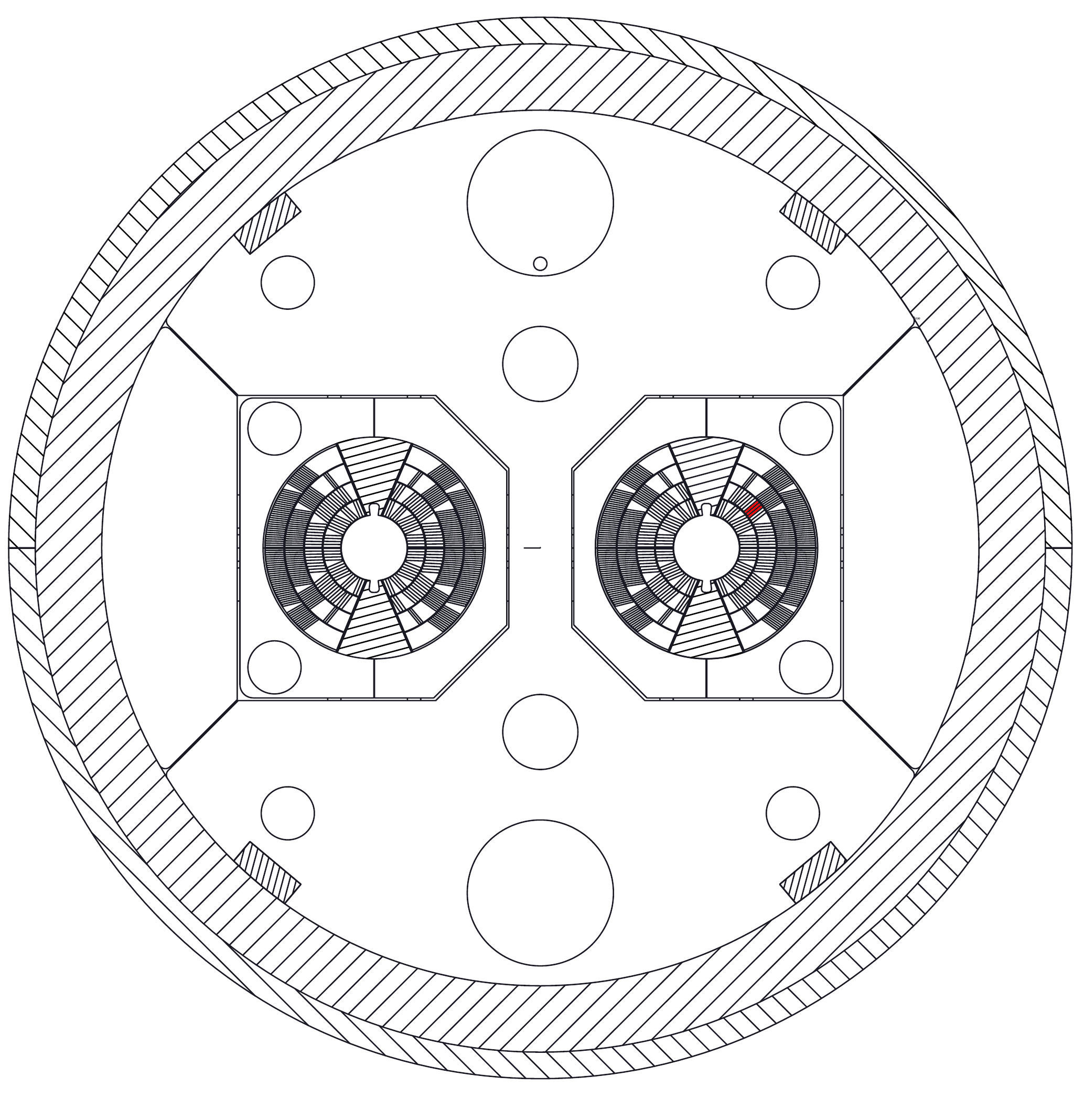}};
		\spy [TUDa-2b] on ($(fig.south west) + \xcord*(fig.south east) - \xcord*(fig.south west) + \ycord*(fig.north west) - \ycord*(fig.south west)$) in node [right] at ($(fig.east) + (1em,0)$);
	\end{tikzpicture}
	\fi%
	\caption{Cross-section of cos-theta magnet for \ac{fcc} and \ac{felhc}. A stack of three Rutherford cables is highlighted in red \cite{Schoerling.2019a}.}
	\label{fig:benchmarkmodel}
\end{figure}

This paper is structured as follows. 
First, an introduction to wavelets is given and a spectral method using wavelets is derived. 
Then, the \ac{fem} and the spectral wavelet method are combined into a hybrid \ac{q3d} method. 
Next, an adaptive resolution strategy is discussed for the spectral method. 
Lastly, the methods are compared and employed to compute the heat conduction in a benchmark model representing a superconducting Rutherford cable as used in superconducting accelerator magnets \cite{Schoerling.2019,Schoerling.2019a}, see \Fig~\ref{fig:benchmarkmodel}.
Here, the asymmetrical heat excitation generated by a quench is simulated by an artificial heat source. 
To obtain a realistic quench simulation, a magneto-thermal coupling is required, which would necessitate the use of vectorial edge shape functions \cite{Biro_1999aa}. These are more involved than nodal ones and beyond this work's scope \cite{Schnaubelt_2021aa}.

\section{Introduction to wavelets}\label{sec:introduct_wav}
\subsection{General statements}
The \ac{mra} \cite{Mallat.1989} consists of closed spaces $\Vj$ that are nested subspaces of each other, \ie it holds that $\Vj\subset\V{j-1}$. Furthermore, the spaces $\Wj:=\Vj^{\perp}\cap \V{j-1}$ are defined as the orthogonal complement of $\Vj$ in $\V{j-1}$. Thus, it holds that
\begin{equation}\label{eq:direct_sum_small}
	\V{j-1} = \Vj\oplus\Wj
\end{equation}
and
\begin{equation}
	\W{i} \perp \W{j}
\end{equation} 
for $i\neq j$. By applying \eqref{eq:direct_sum_small} to itself, one gets
\begin{equation}
	\Vj = V_J \oplus \bigoplus_{k=0}^{J-j-1}W_{J-k}\,,
\end{equation}
where $\oplus$ denotes the direct sum. There are \emph{scaling functions} $\phi_{j,n}$ and \emph{wavelets} $\psi_{j,n}$. They are equipped with a scale $j$ and a translation $n$ and are defined as 
\begin{alignat}{2}
	\phi_{j,n}(x) &:= 2^{-\nicefrac{j}{2}}\phi(2^{-j}x -n)\,,\\
	\psi_{j,n}(x) &:= 2^{-\nicefrac{j}{2}}\psi(2^{-j}x -n)\,.
\end{alignat}
Within these definitions, the \emph{mother scaling function} $\phi$ and the \emph{mother wavelet} $\psi$ are used. These coincide with the scaling function and wavelet, respectively, for $j=0$ and $n=0$. 

The scaling functions and wavelets at fixed scale $j$ constitute an orthonormal basis to the space $\Vj$ and $\Wj$, respectively. To be more precise, it holds that
\begin{alignat}{2}
	\Vj &:=\overline{\spn\setprop{\phi_{j,n}}{n\in\Z}}\,,\\
	\Wj &:=\overline{\spn\setprop{\psi_{j,n}}{n\in\Z}}\,.\\
\end{alignat}

\subsection{\Acl*{dbn} kind wavelets}\label{sec:dbn}
\begin{figure}
	\centering
	\ifpdf%
	\includegraphics{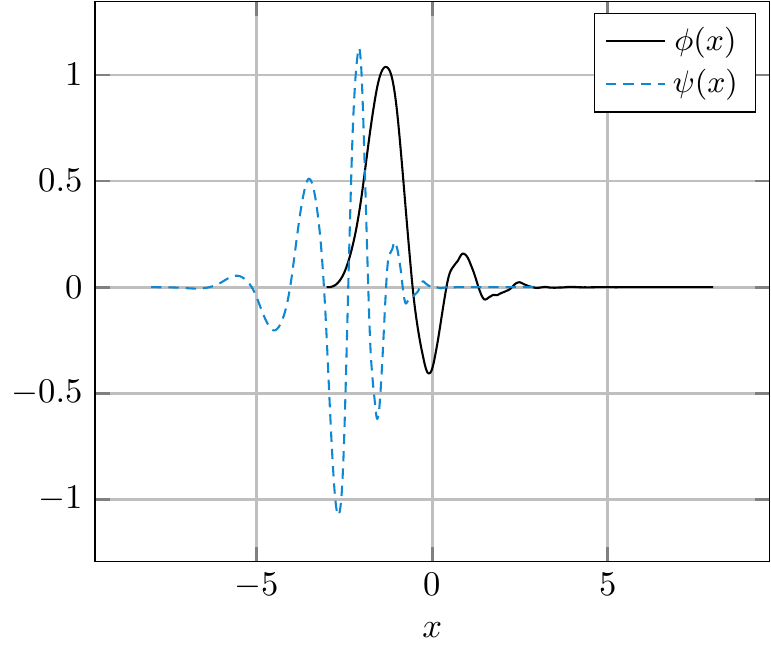}%
	\else%
	\tikzsetnextfilename{scaling_wavelet_N6}%
	\begin{tikzpicture}
		\pgfplotsset{table/search path={plot/data}}
		\begin{axis}[
			xlabel = $x$,
			grid= major,
			cycle list name=myColorCycleList
			]
			\addplot+[wavelet style] table[x expr=\thisrow{x_sf}+2,y=scal] {scaling_wav_N6_j8.dat};\addlegendentry{$\phi(x)$}
			\addplot+[wavelet style,densely dashed] table[x expr=\thisrow{x_wav}+2,y=wave] {scaling_wav_N6_j8.dat};\addlegendentry{$\psi(x)$}
		\end{axis}
	\end{tikzpicture} 
	\fi%
	\caption{Scaling function $\phi(x)$ (black and solid) and wavelet $\psi(x)$ (blue and dashed) of \ac{dbn} kind with $N=6$.}
	\label{fig:wavelet}
\end{figure}
There is a great variety of different kinds of wavelets \cite{Daubechies.1992,Gomes.2015}. However, within the spectral methods of this work, scaling functions and wavelets of the \ac{dbn} kind are used, because they are compactly supported and constructed in a way to have a maximum number of vanishing moments \cite{Daubechies.1992}. To be more precise, the wavelets have a number of $N$ vanishing moments, \ie it holds
\begin{equation}
	\int x^k\psi(x)\d x = 0
\end{equation}
for $k=0,\dots,N-1$. Moreover, they are constructed such that they have a minimal support width for the given number of vanishing moments \cite{Mohlenkamp.2008}. 
The scaling functions used in this work are continuous for $N>1$. In \cite{Rioul.1992} one can find estimates for the Hölder regularity for the scaling functions with $N=2,\dots,10$. For $N=2,3,4$, these are also determined in \cite{Daubechies.1992}. It follows that the scaling functions are continuously differentiable for $N=3,4,5$ and two times continuously differentiable for $N=6,7,8$. The wavelets have the same regularity properties.
For $N=6$, scaling function and wavelet are depicted in \fig~\ref{fig:wavelet}.

The differential equations to be solved in the subsequent sections are formulated on intervals. Consequently, for $N\geq2$ there are always functions that are partially inside and partially outside the interval. But a basis that is defined on the interval is required. The construction of such a basis as proposed in \cite{Cohen.1993} is used in this work. Let $\Int=\left[a,b\right]$ be an interval. For each boundary $N$ boundary scaling functions and boundary wavelets are constructed, such that the number of vanishing moments is preserved. Inside the interval, the ordinary functions are used. 

Because the boundary scaling functions and boundary wavelets can be written as a linear combination of scaling functions, they feature the same regularity.
For a sufficient small scale $j$, the scaling functions on the interval $\Int$ are defined by
\begin{equation}\label{eq:solving_1}
	\phihat_{j,n} := \begin{cases}
		\phia{j,n}\,, \quad & n=0,\dots,N-1\,,\\
		\phi_{j,n}\,,\quad & n=N,\dots,\anzInner+N-1\,,\\
		\phib{j,n}\,,\quad & n= \anzInner+N,\dots,\anz-1\,.
	\end{cases}
\end{equation}
Herein, $\anz$ indicates the number of basis functions on the interval and $\anzInner$ indicates the number of interior basis functions. With this definition, the space of scaling functions at scale $j$ on the interval $\Int$ can be defined as
\begin{equation}
	\VInt{j}:=\setprop{\phihat_{j,n}}{n=0,\dots,\anz-1}\,.
\end{equation}
The space of wavelets at scale $j$ on the interval $\Int$, indicated by $\WInt{j}$, is defined analogously. 

\section{Spectral methods with wavelets-bases}\label{sec:spectral_method}
As stated above, the \ac{q3d} methods is a combination of a \ac{2D} \ac{fem} and a \ac{1D} spectral method. The latter one, that uses a wavelet-basis, is examined in this section. It is exemplary developed for the \ac{1D} heat equation. The \ac{pde} together with the boundary conditions and the initial condition reads as
\begin{equation}\label{eq:heat_prob_1d}
	-\delx\big(\lambda(x)\delx\theta(x,t)\big) + \cv(x)\delt\theta(x,t) = q(x,t)\,,
\end{equation}
with $(x,t)\in\Omegat$,
\begin{alignat}{3}
	\theta(x,t) &= \thetadir(x,t)\,,\quad  &(x,t)&\in\Gammadir\times(0,\infty)\,,\\
	-\lambda(x)\delx\theta(x,t) &= \thetaneum(x,t)\,, \quad  &(x,t)&\in\Gammaneum\times(0,\infty)\,,\\
	\theta(x,0) &= \theta_0(x)\,, \quad &x&\in\Omega\,.
\end{alignat}
Here, $x$ is the spatial coordinate in \si{\meter}, $t$ the time in \si{\second}, $\theta(x,t)$ the unknown temperature in \si{\kelvin}, $\lambda(x)$ the thermal conductivity in \si{\watt\per\meter\per\kelvin}, $\cv(x)$ the volumetric heat capacity in \si{\joule\per\meter\cubed\per\kelvin} and $q(x,t)$ the volumetric heat flux density in \si{\watt\per\meter\cubed}. Furthermore, the problem is defined on the domain $\Omegat:=\Omega\times\left(0,\infty\right)$, where $\Omega$ is the spatial domain. The boundary $\partial\Omega$ is divided in $\Gammadir$ and $\Gammaneum$, with $\partial\Omega = \Gammadir\cup\Gammaneum$ and $\Gammadir\cap\Gammaneum=\emptyset$, on which Dirichlet $\thetadir(x,t)$ in \si{\kelvin} and Neumann boundary conditions $\thetaneum(x,t)$ in \si{\watt\per\meter\squared} are defined, respectively. The initial condition uses the initial value $\theta_0(x)$ in \si{\kelvin}. 

Let for this section be $\Omega=(0,\L)$. The temperature $\theta(x,t)$ is then approximated by $\thetatil(x,t)$ in $\VOmega{j}$ for each time instance. The ansatz of method of lines leads to
\begin{equation}\label{eq:approx_theta}
	\theta(x,t)\approx\thetatil(x,t) = \sum_{n=0}^{\anz-1}u_n(t)\phihat_{j,n}(x)\,.
\end{equation}
The Galerkin approach is applied to \eqref{eq:heat_prob_1d}. Therefore, \eqref{eq:heat_prob_1d} is tested with the basis functions of $\VOmega{j}$. With the inner product $\innerprod{u(x)}{v(x)}:=\int_\Omega u(x)v(x)\d x$, this leads to
\begin{equation}\label{eq:tested}
	\innerprod{-\delx\big(\lambda(x)\delx\theta(x,t)\big)}{\phihat_{j,m}(x)} + \innerprod{\cv(x)\delt\theta(x,t)}{\phihat_{j,m}(x)} = \innerprod{q(x,t)}{\phihat_{j,m}(x)},
\end{equation}
for $m = 0,\dots,\anz-1$. After inserting \eqref{eq:approx_theta} into \eqref{eq:tested}, one receives the semi-discrete equation
\begin{equation}\label{eq:discrete_form}
	\Asf\u(t) + \Msfcv\delt\u(t) = \qsf(t)\,.
\end{equation}
The used components are defined as 
\begin{alignat}{2}
	\Asf&:=\left(\Asfe[;\,m,n]\right)_{m,n=0,\dots,\anz-1}\in\R^{\anz\times\anz}\,,\\
	\Msfcv&:= \left(\Msfecv[;\,m,n]\right)_{m,n=0,\dots,\anz-1}\in\R^{\anz\times\anz}\,,\\
	\qsf(t)&:= \left(\qsfe_m\left(t\right)\right)_{m=0,\dots,\anz-1}\in\R^\anz\,,\\
	\u(t)&:=\left(u_n\left(t\right)\right)_{n=0,\dots,\anz-1}\in\R^{\anz}\,,
\end{alignat}
where
\begin{alignat}{2}
	\Asfe[;\,m,n]&:=\int_\Omega \lambda(x)\delx\phihat_{j,n}(x)\delx\phihat_{j,m}(x)\d x -\left[\lambda(x)\delx\phihat_{j,n}(x)\phihat_{j,m}(x)\right]_0^\L\,,\\
	\Msfecv[;\,m,n]&:=\int_\Omega \cv(x)\phihat_{j,n}(x) \phihat_{j,m}(x) \d x\,,\\
	\qsfe_m(t)&:=\int_\Omega q(x,t)\phihat_{j,m}(x) \d x\,.
\end{alignat}
Integration by parts is used in the calculation of the entries of $\Asf$. For homogeneous material properties, $\Msfcv$ is a diagonal matrix due to the orthonormality of the scaling functions. Consequently, $\Msfcv$ is a regular matrix and \eqref{eq:discrete_form} is a system of ordinary differential equations. 

For the boundary conditions to be satisfied, one can find a matrix $\Bsf\in\R^{2\times\anz}$ and a vector $\bsf\in\R^{2\times1}$ such that the equation 
\begin{equation}\label{eq:disc_bound_cond}
	\Bsf\u(t) = \bsf
\end{equation}
has to hold for every time instance. For discretizing \eqref{eq:discrete_form} in time, an arbitrary time integration method can be used.

\section{\Acl{q3d} method}\label{sec:q3d}
\subsection{Definition of the domain and spaces}
The \ac{q3d} method shall be derived by means of the \ac{3D} heat equation
\begin{equation}\label{eq:heat_eq_3d}
	-\nabla\cdot\big(\lambda(\x)\nabla\theta(\x,t)\big) + \cv(\x)\delt\theta(\x,t) = q(\x,t)\,,
\end{equation}
where $(\x,t)\in\Omegat$. With a \ac{2D} and open domain $\OmegaS\subset\R^2$ and $\Int=(0,\L)$, the spatial domain defined as
\begin{equation}\label{eq:def_prb_3d}
	\Omega := \setprop{\x\in\R^3}{(x,y)\in\OmegaS,\,z\in\Int}
\end{equation}
extends in $z$-direction from 0 to $\L$. Moreover, it it assumed that the geometry is translationally invariant with respect to the $z$-direction. The boundary $\partial\Omega$ is divided in three different parts
\begin{subequations}\label{eq:def_areas}
	\noeqref{eq:Sfront,eq:Sback,eq:Sside}
	\begin{alignat}{2}
		\Sfront &:= \setprop{\x\in\R^3}{(x,y)\in\OmegaS\,,z=0},\,\label{eq:Sfront}\\
		\Sback &:= \setprop{\x\in\R^3}{(x,y)\in\OmegaS\,,z=\L},\,\label{eq:Sback}\\
		\Sside &:= \partial\Omega\setminus\left\{\Sfront\cup\Sback\right\}\,.\label{eq:Sside}
	\end{alignat}
\end{subequations}

Next the related spaces need to be defined. Let 
\begin{equation}
	\Sm := \setprop{v\in C(\bar{\Omega})}{v|_T \in\Pi_m ,\;\forall T\in\Tcal}
\end{equation}
be the space of polynomial finite elements of order $m$, where $\Pi_m$ is the set of all polynomials up to order $m$ and $\Tcal$ is a two-dimensional triangulation (see \cite{braess.2007}). With this,
\begin{equation}
	\K^j:= \{v\in C(\bar{\Omega}):v(x,y,z) = f(x,y)g(z),\,f\in\Seins,\,g\in\VInt{j}\}\,.
\end{equation}
Combining the bases of both spaces, $\Seins$ and $\VInt{j}$, a basis for $\K^j$ is given by
\begin{equation}
	\setprop{\hat{\bfges}_{j,m,n}=\bffe_m \phihat_{j,n-1}}{m=1,\dots,\anzfe,\,n=1,\dots,\anz}\,,
\end{equation}
with $\bffe_m\in\Seins$, $\phihat_{j,n-1}\in\VInt{j}$ and $\anzfe$ being the number of nodes in the triangulation $\Tcal$. To make the following derivation clearer, the notation
\begin{equation}\label{eq:def_kappa}
	\bfges_{j,k} := \hat{\bfges}_{j,m,n}\,,\qquad k = n+(m-1)\anzfe
\end{equation}
is used to reduce the number of indices from 3 to 2. Thus, the basis for $\K^j$ can be written as
\begin{equation}
	\setprop{\bfges_{j,k}}{k=1,\dots,\anz\anzfe}\,.
\end{equation}

\subsection{Derivation of the \ac{q3d} method}
The \ac{q3d} method is derived for the \ac{3D} heat equation \eqref{eq:heat_eq_3d}. The temperature is approximated in $\K^j$. Using the ansatz of method of lines leads to
\begin{equation}\label{eq:ansatz_3d}
	\theta(\x,t)\approx\thetatil(\x,t) = \sum_{n=1}^{\anz\anzfe}u_n(t)\bfges_{j,n}(\x)\,.
\end{equation}
The heat equation is tested with $\bfges_{j,m}$ for $m=1,\dots,\anz\anzfe$. Together with \eqref{eq:ansatz_3d}, one receives the semi-discrete equation
\begin{equation}\label{eq:discrete_eq_3d}
	\Aqu\u(t) + \Mqu\delt\u(t) = \qqu(t)
\end{equation}
similar to \eqref{eq:discrete_form}. It is assumed that the material parameters can be written as $\lambda(\x) = \lambxy(x,y)\lambz(z)$ and $\cv(\x) = \cvxy(x,y)\cvz(z)$. Then, the matrices can be written as
\begin{align}
	\Aqu &= \Afe\otimes\Msflambda + \Mfelambda\otimes\Asf\label{eq:q3d_discrete}\\
	\intertext{and}
	\Mqu &= \Mfecv\otimes\Msfcv\label{eq:mass_3d}\,,
\end{align}
where $\otimes$ denotes the Kronecker product. The remaining matrices, \ie the stiffness matrix of the \ac{2D} \ac{fem} and mass matrices of the spectral and \ac{fem}, are defined as
\begin{alignat}{2}
	\Afe &:= \left(\Afee[;\,m,n]\right)_{m,n=1,\dots,\anzfe}\in\R^{\anzfe\times\anzfe}\,,\\
	\Mfecv &:= \left(\Mfeecv[;\,m,n]\right)_{m,n=1,\dots,\anzfe}\in\R^{\anzfe\times\anzfe}\,,\\
	\Mfelambda &:= \left(\Mfeelambda[;\,m,n]\right)_{m,n=1,\dots,\anzfe}\in\R^{\anzfe\times\anzfe}\,,\\
	\Msflambda &:= \left(\Msfelambda[;\,m,n]\right)_{m,n=0,\dots,\anz-1}\in\R^{\anz\times\anz}\,,
\end{alignat}
with
\begin{alignat}{2}
	\Afee[;\,m,n] &:=\int_{\OmegaS}\lambxy(x,y)\nabla\bffe_m(x,y)\cdot\nabla\bffe_n(x,y)\d A\,,\\
	\Mfeecv[;\,m,n] &:=\int_{\OmegaS}\cvxy(x,y)\bffe_m(x,y)\bffe_n(x,y)\d A\,,\\
	\Mfeelambda[;\,m,n] &:=\int_{\OmegaS}\lambxy(x,y)\bffe_m(x,y)\bffe_n(x,y)\d A\,,\\
	\Msfelambda[;\,m,n]&:=\int_\Int \lambz(z)\phihat_{j,n}(z) \phihat_{j,m}(z) \d z\,.
\end{alignat}

For homogeneous material properties, $\Mfecv$ is symmetric and positive-definite because of
\begin{equation}
	\xivec^\trans\Mfecv\xivec = \sum_{m,n=1}^{\anzfe}\xi_m\xi_n\Mfeecv[;\,m,n] = \cvxy\int_{\OmegaS}\abs{\sum_{m=1}^{\anzfe}\xi_m\bffe_m(x,y)}^2\d A > 0\,,
\end{equation}
for $\xivec \in\R^{\anzfe}\setminus\left\{0\right\}$. Thus, $\Mfecv$ is a regular matrix. Because $\Msfcv$ is regular as well, $\Mqu$ is non-singular due to the properties of the Kronecker product and thus, \eqref{eq:discrete_eq_3d} is a system of ordinary differential equations.

Under the assumption that the volumetric heat flux density can be written as $q(\x,t) = q_{xy}(x,y,t)q_z(z,t)$, the right hand side of \eqref{eq:discrete_eq_3d} reads as
\begin{equation}
	\qqu(t) = \sfe(t)\otimes\ssf(t)\,,
\end{equation}
with
\begin{align}
	\sfe(t)&:=\left(\sfee_m(t)\right)_{m=1,\dots,\anzfe}\in\R^{\anzfe}\,,\\
	\ssf(t)&:=\left(\ssfe_m(t)\right)_{m=1,\dots,\anz}\in\R^{\anz}
\end{align}
and
\begin{align}
	\sfee_m(t)&:=\int_{\OmegaS}q_{xy}(x,y,t)\bffe_m(x,y) \d A\,,\\ 
	\ssfe_{m}(t)&:=\int_\Int q_z(z,t)\phihat_{j,m}(z) \d z\,.
\end{align}

For the realization of the boundary conditions, the three different parts of the boundary, defined in \eqref{eq:def_areas}, are considered individually. At the front and back side, $\Sfront$ and $\Sback$, this is done with the spectral method, and at the hull $\Sside$ with the \ac{fem}.

\section{Adaptive resolution}\label{sec:adapt_res}
In the spectral method in \sect~\ref{sec:spectral_method}, scaling functions at the same scale are used for the whole interval as basis functions. Thus, the resolution is constant for the whole interval as well. 
When considering a very long domain in $z$-direction, a high resolution becomes computationally very expensive due to the large number of degrees of freedom. 
With the \ac{arm} derived in this section, a resolution depending on the location is chosen utilizing the wavelet transform, such that the overall accuracy is sufficiently high. 
This can strongly decrease the required degrees of freedom. 

\begin{figure}%
	\centering
	\ifpdf%
	\includegraphics{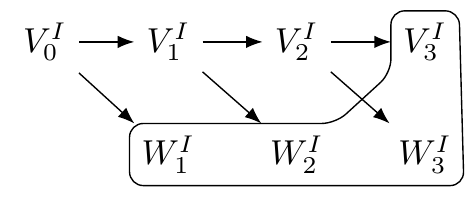}%
	\else%
	\tikzsetnextfilename{space_decomposition}
	\begin{tikzpicture}[>=Latex]
		\matrix (m) [matrix of math nodes,nodes in empty cells,column sep={1.5em},row sep={1.5em}]{
			\VInt{0} & \VInt{1} & \VInt{2} & \VInt{3}\\
			& \WInt{1} & \WInt{2} & \WInt{3}\\
		};
		\draw[->] (m-1-1) -- (m-1-2);
		\draw[->] (m-1-1) -- (m-2-2);
		\draw[->] (m-1-2) -- (m-1-3);
		\draw[->] (m-1-2) -- (m-2-3);
		\draw[->] (m-1-3) -- (m-1-4);
		\draw[->] (m-1-3) -- (m-2-4);
		\draw[rounded corners] (m-2-2.south west) -- (m-2-4.south east) -- (m-1-4.north east) -- (m-1-4.north west) -- (m-1-4.south west) -- (m-2-3.north east) -- (m-2-2.north west) -- cycle;	
	\end{tikzpicture}
	\fi%
	\caption{Decomposition of $\VInt{0}$ in spaces of scaling functions and wavelets of greater scale. The arrows denote a decomposition. The addition of approximations of a function in the boxed spaces yields the same result as approximating the function in $\VInt{0}$. Here, $\maxscale=3$ and $\minscale=1$. }
	\label{fig:wavelet_transform}
\end{figure}%
The boundary basis functions (\sect~\ref{sec:dbn}) are chosen such that the main properties of wavelets still hold. In particular, it holds that $\VInt{j}\subset\VInt{j-1}$, $\WInt{j}={\VInt{j}}^{\perp}\cap \VInt{j-1}$ and 
\begin{equation}\label{eq:direct_sum_interval}
	\VInt{j-1} = \VInt{j}\oplus\WInt{j}\,.
\end{equation}
This allows to decompose the spaces $\VInt{j}$. \Fig~\ref{fig:wavelet_transform} shows an example of such a decomposition.
The space $\VInt{0}$ is decomposed in $\VInt{1}$ and $\WInt{1}$, $\VInt{1}$ in $\VInt{2}$ and $\WInt{2}$, and so on. 
Approximating a function in $\VInt{0}$ is thus equivalent to approximating the same function in the boxed spaces and adding the solutions. 

Let $\maxscale$ and $\minscale$ denote the maximum and minimum scale used in the approximation, respectively. The approximation is then equivalent to one with scaling functions at scale $j=\minscale-1$. The basis for the adaptive resolution is
\begin{align}
	\dynbasis^{\maxscale}_{\minscale} &:=\setprop{\phihat_{\maxscale,k}}{k = 0,\dots,\anz_{\maxscale}-1} \\
	&\cup \left\{\bigcup_{j=\minscale}^{\maxscale}\setprop{\psihat_{j,\Khat_{j,n}-1}}{n=1,\dots,\eta_j}\right\}\,.
\end{align}
It contains all basis functions of $\VInt{\maxscale}$ (first set) to ensure that every point in the interval $\Int$ is covered by at least one function. Moreover, it contains subsets of the bases of the spaces $\WInt{j}$ for $j=\minscale,\dots,\maxscale$ (second set). The information which of the latter are used is stored in the $\eta_j$-tuples $\Khat_j=\left(\Khat_{j,1},\dots,\Khat_{j,\eta_j}\right)$. For these, $\Khat_{j,i}<\Khat_{j,i'}$ for $i<i'$ and $1\leq\Khat_{j,i}\leq\anz_j$ for all $i$. How to choose the tuples is discussed later on page~\pageref{tuples}. 

The tuples are also used to define the truncation matrices
\begin{equation}
	\T_j = \left(\evec_{\Khat_{j,1}}\;\cdots\;\evec_{\Khat_{j,\eta_j}}\right)^\trans\in\R^{\eta_j\times\anz_j}\,,
\end{equation}
where $\evec_{\Khat_{j,n}}\in\R^{\anz_j}$ is the $\Khat_{j,n}$-th unit column vector. The basis $\dynbasis^{\maxscale}_{\minscale}$ is used within a spectral method for the \ac{1D} heat equation from \eqref{eq:heat_prob_1d}. This yields the semi-discrete equation
\begin{equation}
	\Adyn\udyn(t) + \Mdyn\delt\udyn(t) = \qdyn(t)\,.
\end{equation}
Herein, the used vectors and matrices as well as $\Bdyn$, the matrix used for the boundary conditions, are determined with the help of $\As$, $\Ms$, $\Bs$, $\us(t)$ and $\qs(t)$. The subscript s stands for a static resolution, whereas d stands for a dynamic resolution. One would receive the static matrices and vectors if all wavelet basis functions in $\dynbasis^{\maxscale}_{\minscale}$ were used, \ie with $\Khat_j=\left(1,\dots,\anz_j\right)$ for $j=\minscale,\dots,\maxscale$. Because of \eqref{eq:direct_sum_interval}, these matrices and vectors can be derived from those from \sect~\ref{sec:spectral_method} at scale $\minscale-1$. With the transformation matrix
\begin{equation}
	\T^{\maxscale}_{\minscale} = \begin{pmatrix}
		\T_{\minscale}&  &  &  \\
		& \ddots &  &  \\[3pt]
		&  & \T_{\maxscale} &  \\
		&  &  & \I_{\anz_{\maxscale}}\\
	\end{pmatrix}\in\R^{\eta\times\anz_{\minscale-1}}\,,
\end{equation}
where $\I_{\anz_{\maxscale}}\in\R^{\anz_{\maxscale}\times\anz_{\maxscale}}$ is the identity matrix and $\eta:=\anz_{\maxscale}+\sum_{j=\minscale}^{\maxscale}\eta_j$, one obtains 
\begin{alignat}{2}
	\Adyn&=\T^{\maxscale}_{\minscale}\As\left(\T^{\maxscale}_{\minscale}\right)^\trans\,,\label{eq:Ad_to_As}\\
	\Mdyn&=\T^{\maxscale}_{\minscale}\Ms\left(\T^{\maxscale}_{\minscale}\right)^\trans\,,\label{eq:Md_to_Ms}\\
	\Bdyn&= \Bs\left(\T^{\maxscale}_{\minscale}\right)^\trans \,,\label{eq:Bd_to_Bs}\\
	\qdyn(t)&=\T^{\maxscale}_{\minscale}\qs(t)\,,\\
	\udyn(t)&= \T^{\maxscale}_{\minscale}\us(t)\,.
\end{alignat}
When changing the resolution, the matrices $\Adyn$ and $\Mdyn$ have to be recalculated. Because $\As$ and $\Ms$ are known beforehand and $\T^{\maxscale}_{\minscale}$ is sparse, the computational costs for recalculation are low. 

\label{tuples}It remains to determine the tuples $\Khat_j$, where $\scalarp{\cdot}{\cdot}:=\abs{\innerprod{\cdot}{\cdot}}$ is used. Note that $\scalarp{\cdot}{\cdot}$ is not an inner product but an abbreviation. The initial condition $\theta_0(x)$ is used at the beginning of the simulation: 
\begin{equation}
	k+1\in\Khat_j\;\Leftrightarrow\;\scalarp{\theta_0(x)}{\psihat_{j,k}}>\tolu\,.
\end{equation}
The basis function $\psihat_{j,k}$ is used if the corresponding coefficient is greater than a tolerance $\tolu>0$,~\ie if the basis function $\psihat_{j,k}$ has a significant contribution to the solution. 

\begin{figure}
	\centering
	\newcommand{\bottomcolor}{TUDa-2b}
	\ifpdf%
	\includegraphics{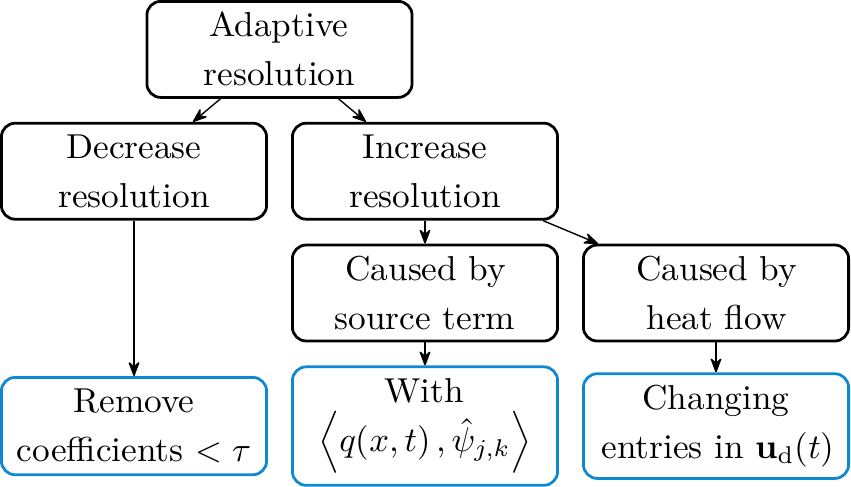}%
	\else%
	\tikzsetnextfilename{flowchart}%
	\begin{tikzpicture}[>={Stealth[round]},block/.style ={rectangle, draw=black, thick,
			text width=7em,align=center, rounded corners,
			minimum height=2em}]
		\graph[layered layout, nodes={block}]{Adaptive resolution ->{ Decrease resolution -> {{"Remove coefficients $<\tolu$" [> minimum layers=2,draw=\bottomcolor]}} , Increase resolution -> {Caused by source term -> {"With $\scalarp{q(x,t)}{\psihat_{j,k}}$" [draw=\bottomcolor]} , Caused by heat flow -> {"Changing entries in $\udyn(t)$" [draw=\bottomcolor]}}}};
	\end{tikzpicture}
	\fi%
	\caption{Flow chart for adapting the resolution.}
	\label{fig:resolution}
\end{figure}

During the simulation it may occur that the resolution needs to be adapted. One has to distinguish between decreasing and increasing the resolution (see \Fig~\ref{fig:resolution}). For the former case, after every time step those basis functions whose coefficients are smaller than $\tolu$ are removed from $\dynbasis^{\maxscale}_{\minscale}$. For the latter case, two different causes are considered. An increase of the resolution may be necessary due to the source term $q(x,t)$ or heat flow. Because the source term is known beforehand, the required resolution can be determined as for the initial condition. The function $\psihat_{j,k}$ is used if $\scalarp{q(x,t)}{\psihat_{j,k}}>\tolu$. 

Because the solution of the \ac{pde} is approximated with the basis $\dynbasis^{\maxscale}_{\minscale}$, it is challenging to detect where a higher resolution caused by heat flow is needed. One possible approach is presented in the following. 

Whenever a coefficient in $\udyn(t)$ changes from one time step to another by a certain factor $\fact$, the resolution in the interval of the support of the corresponding basis function $\psihat_{j,k}$ is increased. Therefore, the wavelets at scale $j-1$, whose support is a subset of the support of $\psihat_{j,k}$, are added to $\dynbasis^{\maxscale}_{\minscale}$, if $j-1$ does not exceed the minimum scale $\minscale$, \ie~if $j>\minscale$. Assuming that $j>\minscale$, the scaling functions at scale $j-1$ with the indices
\begin{equation}
	\Lambda := \left\{-N+1+2k,\dots,N+2k\right\}\cap\left\{0,1,\dots,L2^{-j+1}-1\right\}
\end{equation}
are added to $\dynbasis^{\maxscale}_{\minscale}$, \ie~$\tilde{k}+1\in\Khat_{j-1}$ for all $\tilde{k}\in\Lambda$. The corresponding coefficients to these wavelets are initialized with 0. Hence, the added wavelets must not be removed from $\dynbasis^{\maxscale}_{\minscale}$ for a time interval of length $\numkeep$ 
because the coefficients may need a certain time to exceed the tolerance $\tolu$. This approach has been implemented and is used for the results in \sect~\ref{sec:results_arm}.

\section{Adaptive \Acl{q3d} method}\label{sec:adaptove_q3d}
The \ac{arm} from the previous section is combined with a \ac{2D} \ac{fem} into an \ac{aq3d} method. The adaptivity only takes place in the longitudinal direction, \ie the \ac{2D} mesh in the cross-section does not change. The method is derived for the \ac{3D} heat equation \eqref{eq:heat_eq_3d} and on the domain $\Omega$ defined in \eqref{eq:def_prb_3d}. The space for basis and test functions is 
\begin{equation}
	\K^{\maxscale}_{\minscale}:= \{v\in C(\bar{\Omega}):v(x,y,z) = f(x,y)g(z),\,f\in\Seins,\,g\in\dynbasis^{\maxscale}_{\minscale}\}\,.
\end{equation}
In contrast to the \ac{q3d} method, here the function $g$ is chosen from the space $\dynbasis^{\maxscale}_{\minscale}$. 
With $\setprop{\chi_n}{n = 1,\dots,\eta}$ being a basis of $\dynbasis^{\maxscale}_{\minscale}$, a basis for $\K^{\maxscale}_{\minscale}$ is given by
\begin{equation}
	\setprop{\hat{\bfges}_{m,n}=\bffe_m \chi_{n}}{m=1,\dots,\anzfe,\,n=1,\dots,\eta}\,,
\end{equation}
with $\bffe_m\in\Seins$ and $\chi_{n}\in\dynbasis^{\maxscale}_{\minscale}$. 

Similar to \eqref{eq:discrete_eq_3d}, the semi-discrete equation is 
\begin{equation}\label{eq:discrete_eq_arm_3d}
	\Aaqu\u(t) + \Maqu\delt\u(t) = \qaqu(t)\,.
\end{equation}
Herein, the matrices can be written as
\begin{align}
	\Aaqu &= \Afe\otimes\Mdlambda + \Mfelambda\otimes\Adlambda\label{eq:aq3d_discrete}\\
	\intertext{and}
	\Maqu &= \Mfecv\otimes\Mdcv\label{eq:mass_aq3d}\,.
\end{align}
With \eqref{eq:Ad_to_As} and \eqref{eq:Md_to_Ms} and the definition
\begin{equation}
	\Q^{\maxscale}_{\minscale} := \I_{\anzfe}\otimes \T^{\maxscale}_{\minscale}\,,
\end{equation}
where $\I_{\anzfe}\in \R^{\anzfe\times\anzfe}$ is the identity matrix, the matrices can be written as
\begin{align}
	\Aaqu &= \Q^{\maxscale}_{\minscale}\left(\Afe\otimes\Mslambda + \Mfelambda\otimes\Aslambda\right)\left(\Q^{\maxscale}_{\minscale}\right)^\trans\\
	\intertext{and}
	\Maqu &= \Q^{\maxscale}_{\minscale}\left(\Mfecv\otimes\Mscv\right)\left(\Q^{\maxscale}_{\minscale}\right)^\trans\,.
\end{align}
The middle term is independent of the resolution and has to be calculated only once at the beginning of the simulation. 

Similar to the matrices, the right-hand side vector is
\begin{equation}
	\qaqu(t) = \Q^{\maxscale}_{\minscale} \left(\sfe(t)\otimes\qs(t)\right)\,.
\end{equation}
The boundary conditions are realized at $\Sfront$ and $\Sback$ with the \ac{arm} and at $\Sside$ with the \ac{fem}. 
The process of changing the resolution is very similar to the \ac{arm}. Here, a separate resolution is calculated for every edge in $z$-direction in the same way as explained in \sect~\ref{sec:adapt_res}. The resolutions of all edges and the source term are then merged with each other to get a single resolution for all edges. Consequently, the matrix $\Q^{\maxscale}_{\minscale}$ can be recalculated and the simulation can proceed. 

\section{Simulation results}\label{sec:simul_results}
The time discretization for each method is realized with the implicit Euler method, where $\Delta t$ denotes the step width.
The difference between numerical and analytical solution is measured with \begin{align}
	\maxnorm{f} &= \max_i\left\{\max_{x\in\Omega}\abs{f(x,t_i)}\right\}\\
	\intertext{and}
	\norm{\x}_\infty &= \max_i\left\{\max_m\abs{x_m(t_i)}\right\}\,,
\end{align}
where $t_i=i\Delta t$ is the time step. The norm $\maxnorm{\cdot}$ is used to compare $\thetatil$ with an analytical solution as a function, whereas $\norm{\cdot}_\infty$ compares the vector $\u$ with a vector $\u_{\mathrm{ana}}$ that contains the coefficients of the analytical solution. 

\subsection{\Acl*{ssm}}
\begin{figure}
	\centering
	\ifpdf%
	\includegraphics{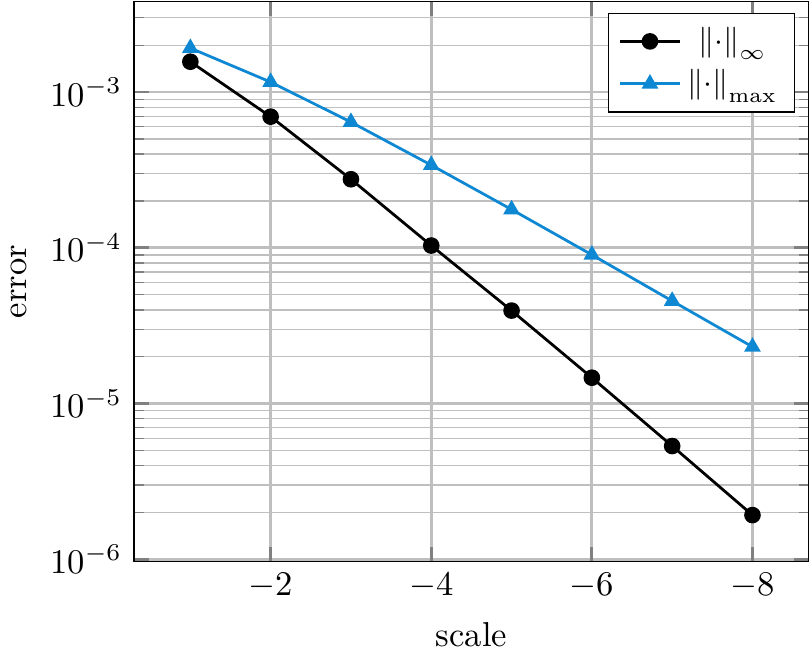}%
	\else%
	\tikzsetnextfilename{konvergence_N6}%
	\begin{tikzpicture}
		\pgfplotsset{table/search path={plot/data}}
		\begin{semilogyaxis}[xlabel={scale},x dir=reverse,
			ylabel={error},
			grid= both,
			cycle list name=myColorCycleList,
			legend style={legend columns=1,at={(.98,.98)},anchor=north east},
			]
			\addplot+[konv style] table[x = scales, y = Inf] {konvergenz_dir_N6.dat};\addlegendentry{$\norm{\cdot}_\infty$}
			\addplot+[konv style,mark=triangle*] table[x = scales, y = max] {konvergenz_dir_N6.dat};\addlegendentry{$\maxnorm{\cdot}$}
		\end{semilogyaxis}
	\end{tikzpicture}
	\fi%
	\caption{Error between the numerical and analytical solution in the $\norm{\cdot}_\infty$ and the $\maxnorm{\cdot}$ norm.}
	\label{fig:konv_dir_N6}
\end{figure}

For the \ac{ssm}, the \ac{1D} heat equation \eqref{eq:heat_prob_1d} with dimensionless $\L=10$, $N=6$, $\lambda=10$, $\cv=1$, $\Delta t = \num{1e-3}$ and 100 time steps is solved. The initial condition is $	\theta(x,0) = \sin\left(\frac{\pi}{\L}x\right)$, the source term is $q(x,t)=0$, and homogeneous Dirichlet boundary conditions are applied. 
The error between numerical and analytical solution is shown in \fig~\ref{fig:konv_dir_N6}. As expected, the error in both norms decreases with decreasing scale.

\subsection{\Acl{q3d} method}\label{sec:results_q3d}
The \ac{q3d} method is used to simulate the temperature distribution for the benchmark problem defined in \cite{DAngelo.2020} and highlighted in red in \Fig~\ref{fig:benchmarkmodel}. The domain consists of three insulated cables. They extend from $z=\SI{0}{\meter}$ to $z=\SI{1}{\meter}$ and are surrounded and separated by an insulation material (glass fiber). The problem's spatial dimensions and material data are summarized in table \ref{tab:geo_data} and \ref{tab:mat_data}, respectively. In the model, the cables have been homogenized into a bulk model consisting of one material, although in reality they are built up of several strands \cite{Rochepault.2016}.
\begin{table}%
	\centering
	\small
	\renewcommand{\arraystretch}{1.2}
	\caption{Geometrical data of the stack of Rutherford cables (taken from \cite{DAngelo.2020}).}
	\begin{tabular}{|l|c||l|c|}
		\hline
		Cable width: & \SI{1.5}{\milli\meter}&Total width: & \SI{4.9}{\milli\meter}\\\hline
		Cable height: & \SI{15}{\milli\meter}&Total height: & \SI{15.2}{\milli\meter}\\\hline
		Insulation thickness: & \SI{0.1}{\milli\meter}&Model length: & \SI{1}{\meter}\\\hline
	\end{tabular}
	\label{tab:geo_data}
	\renewcommand{\arraystretch}{1}
\end{table}%
\begin{table}%
	\centering
	\small
	\renewcommand{\arraystretch}{1.2}
	\caption{Material data of the stack of Rutherford cables \cite{DAngelo.2020}.}
	\begin{tabular}{|l|l|l|}
		\hline
		\multirow{2}{*}{\shortstack[l]{Thermal\\conductivity}} & Cable: & \SI{235.6}{\watt\per\meter\per\kelvin}\\\cline{2-3}
		& Glass fiber: & \SI{0.1}{\watt\per\meter\per\kelvin}\\\hline
		\multirow{2}{*}{\shortstack[l]{Volumetric\\heat capacity}}&Cable: & \SI{314.1}{\joule\per\meter\cubed\per\kelvin}\\\cline{2-3}
		& Glass fiber: & \SI{750}{\joule\per\meter\cubed\per\kelvin}\\\hline
	\end{tabular}
	\label{tab:mat_data}
	\renewcommand{\arraystretch}{1}
\end{table}%
An initial condition of $\theta_0(\x) = \SI{2}{\kelvin}$ together with Dirichlet boundary conditions at the front ($z=\SI{0}{\meter}$) and back side ($z=\SI{1}{\meter}$) and homogeneous Neumann boundary conditions are applied at the hull. In order to simulate a quench in the left cable, the time independent source term 
\begin{equation}\label{eq:rutherford_source}
	q(\x,t) = \qmax\e^{-\frac{\left(z-\zq\right)^2}{\sigma^2}}\chiq(\x)
\end{equation}
is applied, where $\chiq(\x)$ is equal to 1 in the left cable and equal to 0 elsewhere. The remaining values are chosen as $\qmax=\SI{1e6}{\watt\per\meter\cubed}$, $\zq=\SI{0.33}{\meter}$ and $\sigma=\SI{0.05}{\meter}$. For the simulation, scaling functions with $N=6$ at scale \num{-5} are used and \num{200} steps of the implicit Euler method with $\Delta t=\SI{5e-5}{\second}$ are performed. This leads to a total simulation time of \SI{10}{\milli\second}.

\newcommand{\mywidth}{.8\columnwidth}
\newcommand{\myheight}{5cm}
\begin{figure}
	\centering
	\ifpdf%
	\includegraphics{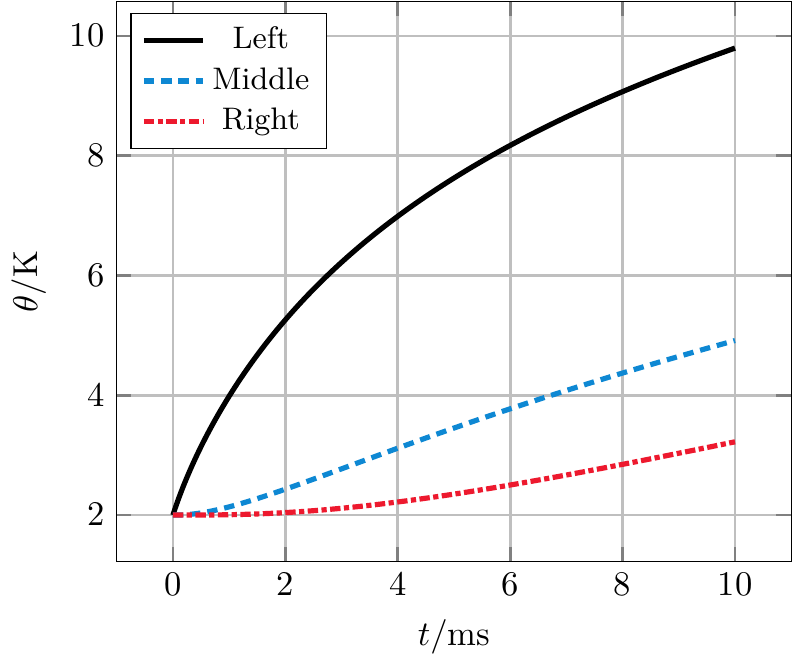}%
	\else%
	\tikzsetnextfilename{temperature_time}%
	\begin{tikzpicture}
		\pgfplotsset{table/search path={plot/data/rutherford}}
		\begin{axis}[xlabel={$t/\si{\milli\second}$},
			ylabel={$\theta/\si{\kelvin}$},
			legend entries={Left,Middle,Right},
			cycle list name=myColorCycleList,
			legend style={legend columns=1,at={(.02,.98)},anchor=north west},
			scaled x ticks = base 10:3,
			xtick scale label code/.code={\empty},
			grid=major,
			]
			\addplot+[normal plot] table[x=time,y=left] {rutherford_N6_temp_knoten.dat};
			\addplot+[normal plot,densely dashed] table[x=time,y=middle] {rutherford_N6_temp_knoten.dat};
			\addplot+[normal plot,densely dashdotted] table[x=time,y=right] {rutherford_N6_temp_knoten.dat};
		\end{axis}
	\end{tikzpicture}
	\fi%
	\caption{Evolution of the temperature in the three cables over time. Evaluated at $z=\zq$ and in the center of the cross-section of each cable.}
	\label{fig:rutherford_temp_t}
\end{figure}%
\begin{figure}
	\centering
	\ifpdf%
	\includegraphics{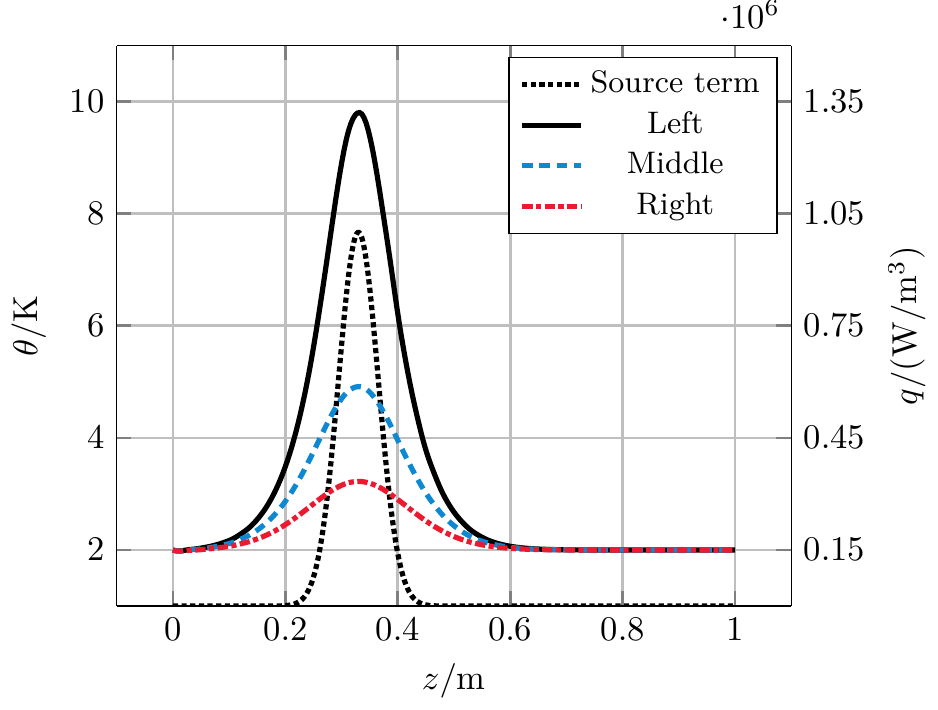}%
	\else%
	\tikzsetnextfilename{temperature_space}%
	\begin{tikzpicture}
		\pgfplotsset{compat=1.10}
		\pgfplotsset{table/search path={plot/data/rutherford}}
		\begin{axis}[ylabel={$q/ (\si{\watt\per\meter\cubed})$},
			xlabel={$z/\si{\meter}$},
			axis y line*=right,
			ymin=0,
			ymax=1.5e6,
			ytick={1.5e5,4.5e5,...,1.5e6},
			grid=major,
			]
			\addplot+[normal plot,black,densely dotted,domain=0:1,samples=800] {1e6*exp((-(x-0.33)^2)/(0.05)^2)};
		\end{axis}
		\begin{axis}[ylabel={$\theta/\si{\kelvin}$},
			ymin=0,
			ymax=1.3e6,
			ymin=1,
			ymax=11,
			ytick={2,4,...,10},
			axis x line=none,
			cycle list name=myColorCycleList,
			legend style={legend columns=1,at={(.98,.98)},anchor=north east},
			]
			\addlegendimage{black,densely dotted}\addlegendentry{Source term}
			\addplot+[normal plot] table[x=x_left,y=y_left] {rutherford_N6_z_dir.dat};\addlegendentry{Left}
			\addplot+[normal plot,densely dashed] table[x=x_mid,y=y_mid] {rutherford_N6_z_dir.dat};\addlegendentry{Middle}
			\addplot+[normal plot,densely dashdotted] table[x=x_right,y=y_right] {rutherford_N6_z_dir.dat};\addlegendentry{Right}
		\end{axis}
	\end{tikzpicture}
	\fi%
	\caption{Temperature distribution in $z$-direction in the three cables at time $t=\SI{10}{\milli\second}$ and the source term in dependency of $z$. The temperature of the cables is evaluated in the center of their cross-section.}
	\label{fig:rutherford_temp_z}
\end{figure}%
\begin{figure}
	\centering
	\ifpdf%
	\includegraphics{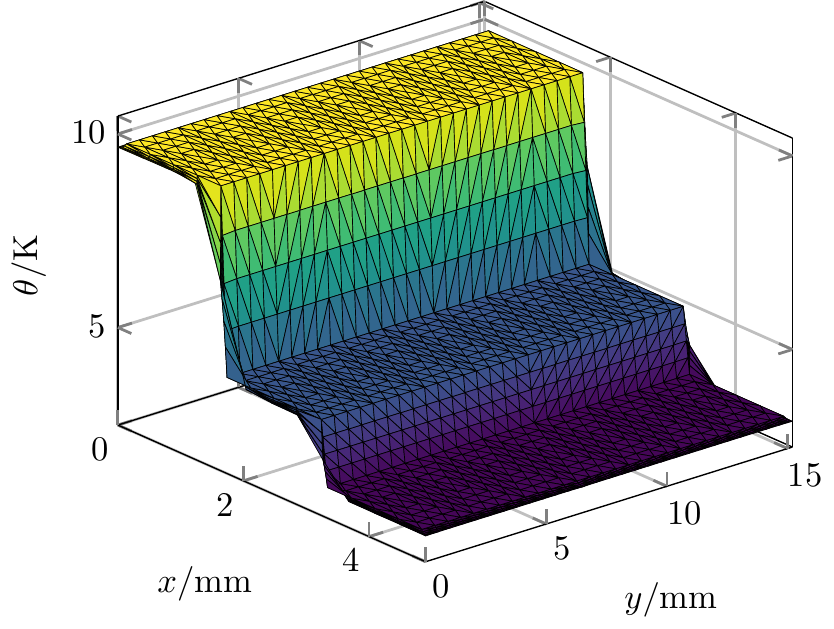}%
	\else%
	\tikzsetnextfilename{temperature_3D}%
	\begin{tikzpicture}
		\pgfplotsset{table/search path={plot/data/rutherford}}
		\begin{axis}[grid=major,
			view={50}{30},
			colormap name=viridis,
			colorbar style ={ytick distance=1},
			xlabel={$x/\si{\milli\meter}$},
			ylabel={$y/\si{\milli\meter}$},
			zlabel={$\theta/\si{\kelvin}$},
			scaled x ticks = base 10:3,
			xtick scale label code/.code={\empty},
			scaled y ticks = base 10:3,
			ytick scale label code/.code={\empty}]
			\addplot3 [patch,very thin,shader=flat,draw=black,z buffer=none] table {alt_mesh.dat};
		\end{axis}
	\end{tikzpicture}
	\fi%
	\caption{Temperature distribution in the Rutherford cable at $z=\zq=\SI{0.33}{\meter}$ at time $t=\SI{10}{\milli\second}$.}
	\label{fig:rutherf_temp_dist}
\end{figure}%
\let\mywidth\relax%
\let\myheight\relax%
In \fig~\ref{fig:rutherford_temp_t}, the evolution of the temperature inside the three cables is shown. For this plot and for \fig~\ref{fig:rutherford_temp_z}, the temperature was evaluated at $z=\zq$ and for each cable in the center of its cross-section. The temperature of the left cable increases immediately, while the other cables start to heat up after a certain time. At the end of the simulation (at $t=\SI{10}{\milli\second}$) the left cable has thus the highest temperature. This can be seen in \fig~\ref{fig:rutherford_temp_z} and \fig~\ref{fig:rutherf_temp_dist} as well. In \fig~\ref{fig:rutherford_temp_z} one can additionally see how the heat has already been flown from $z=\SI{0.33}{\meter}$ in positive and negative $z$-direction. \Fig~\ref{fig:rutherf_temp_dist} shows the temperature distribution at the end of the simulation at $z=\zq$ and displays the mesh in the cross-section. One can see that the temperature inside the three cables is nearly constant, whereas in the insulation material high temperature gradients are present. This temperature distribution is the result of the strongly differing values of the thermal conductivity of the cable and insulation (table \ref{tab:mat_data}) and of the source term in \eqref{eq:rutherford_source}, that is independent of $x$ and $y$ inside the left cable. 

The \ac{q3d} method with scaling functions (denoted with \qdsf) is now compared to a \qdpoly method (from \cite{DAngelo.2020}) and to a conventional \ac{3D} \ac{fem}. The \qdpoly method uses a polynomial basis with a maximum degree $M$ for the spectral method in longitudinal direction. It decomposes the longitudinal direction in non-equidistant intervals called elements. The problem that is solved can be found in \eqref{eq:heat_eq_3d} with dimensionless $L=10$, $\lambda=10$, $\cv=5$, $\Delta t = \num{1e-4}$ and 10 time steps. The initial condition is 
\begin{equation}\label{eq:initial_condition}
	\theta_0(\x) = \cos(\pi x)\cos(\pi y)\sin\left(4\frac{2\pi}{\L} z\right)
\end{equation}
and the source term is $q(\x,t)=0$. To the front and back side homogeneous Dirichlet boundary conditions and to the hull homogeneous Neumann boundary conditions are applied. With the chosen initial condition, the analytical solution of this problem is found by separation of variables as
\begin{equation}
	\theta(\x,t) = \cos(\pi x)\cos(\pi y)\sin\left(\frac{8\pi}{\L} z\right)\e^{-\frac{\lambda}{\cv}\pi^2\left(2+\frac{8^2}{\L^2}\right)t}\,.
\end{equation}

The \qdsf method is used with $N=6$ and the \qdpoly method is used with a maximum polynomial degree $M=4$ on each element in $z$-direction. The \ac{fem} uses a tetrahedral mesh with linear, nodal basis functions. 

\begin{figure}
	\centering
	\ifpdf%
	\includegraphics{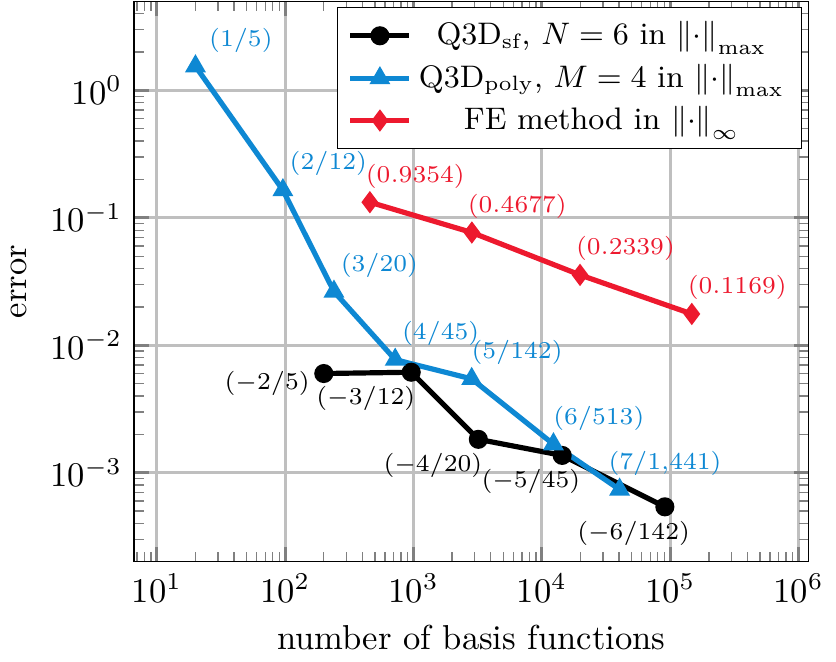}%
	\else%
	\tikzsetnextfilename{comparison}%
	\begin{tikzpicture}
		\pgfplotsset{table/search path={plot/data/comparison}}
		\begin{loglogaxis}[konv axis,grid= major,
			xmax=1.2e6,
			ymin=2e-4,
			ymax=5,
			xlabel={number of basis functions},
			ylabel={error},
			legend style={legend columns=1,at={(.99,.99)},anchor=north east},
			cycle list name=myCycleList,]
			\addplot+[nodes near coords={\scriptsize$(\pgfmathprintnumber[fixed,precision=0] \scale/\pgfmathprintnumber[fixed,precision=0]\anzahl)$},
			visualization depends on={\thisrow{scale} \as \scale},
			visualization depends on={\thisrow{anz_fe}\as \anzahl},
			coordinate style/.condition={\coordindex==0}{anchor=10},
			coordinate style/.condition={\coordindex==1}{anchor=30},
			coordinate style/.condition={\coordindex==2}{anchor=30},
			coordinate style/.condition={\coordindex>2}{anchor=40},
			] table[x=anzahl,y=fehler] {daten_q3d_N6_L10.dat};\addlegendentry{\qdsf, $N=6$ in $\maxnorm{\cdot}$}
			
			\addplot+[nodes near coords={\scriptsize$(\pgfmathprintnumber[fixed,precision=0] \elements/\pgfmathprintnumber[fixed,precision=0]\anzahl)$},
			nodes near coords style={anchor=210},
			visualization depends on={\thisrow{elements} \as \elements},
			visualization depends on={\thisrow{anz_fe}\as \anzahl},
			] table[x=anzahl,y=fehler] {konvergenz_poly_beide_elems_7_N4_L10.dat};\addlegendentry{\qdpoly, $M=4$ in $\maxnorm{\cdot}$}
			\addplot+[
			nodes near coords={\scriptsize$(\pgfmathprintnumber[fixed,precision=4]{\len})$},
			nodes near coords style={anchor=210},
			visualization depends on={\thisrow{len} \as \len},skip coords between index={0}{1},
			] table[x=anzahl,y=fehler] {daten_fe_struct.dat};\addlegendentry{\acs{fem} in $\norm{\cdot}_\infty$}
		\end{loglogaxis}	
	\end{tikzpicture}
	\fi%
	\caption{Error between analytical and numerical solution of \qdsf method (black) and \qdpoly method (blue) and of a conventional \ac{3D} \ac{fem}. 
		The numbers between the paranteses are the scale and the number of FE nodes for \qdsf, the number of elements and the number of FE nodes for \qdpoly and the longest edge length for the \ac{fem}.}
	\label{fig:comparison}
\end{figure}

\Fig~\ref{fig:comparison} shows the error of the three methods with respect to the total number of basis functions. The error for the \ac{q3d} methods are measured in the maximum norm over all edges in $z$-direction and for the \ac{fem} the norm $\norm{\cdot}_\infty$ is used. The computed errors are comparable to each other. Both \ac{q3d} methods lead to significantly smaller errors than the \ac{fem} with a similar number of basis functions. Between the two \ac{q3d} methods there are no major differences. The errors of both methods are similar. However, the number of basis functions in longitudinal direction is more flexible in the \qdpoly method because the polynomial degree and the number of elements can be chosen independently. For the \qdsf method in contrast, once $N$ is set, only the scale can be varied what leads to a fixed number of basis functions. 

\subsection{\Acl*{arm}}\label{sec:results_arm}
The \ac{arm} is used to solve the \ac{1D} heat equation \eqref{eq:heat_prob_1d}. The initial condition 
\begin{equation} 
	\theta_0(x) = \e^{-\frac{\left( x-\nicefrac{\L}{2}\right)^2}{2\sigma^2}}
\end{equation} 
and the boundary conditions $\theta(0,t) = \theta_0(0)\ast\funsol(0,t)$ and $\theta(\L,t) = \theta_0(\L)\ast\funsol(\L,t)$, for $t>0$, are used. Herein, $\funsol(x,t)$ is the fundamental solution of the \ac{1D} heat equation \cite{Liqiu.2008} and $\ast$ denotes convolution. Furthermore, $\L=10$, $N=3$, $\lambda=10$, $\cv=1$, $\sigma=\num{0.15}$, $\Delta t = \num{5e-6}$. Moreover, 100 time steps are carried out and a tolerance $\tolu=\num{1e-8}$ is chosen. 

\begin{figure}
	\centering
	\ifpdf%
	\includegraphics{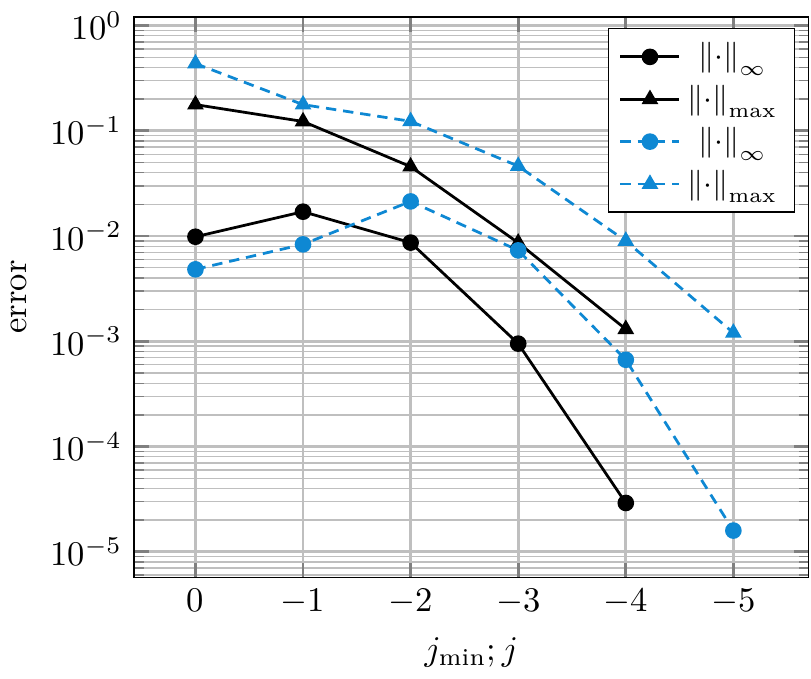}%
	\else%
	\tikzsetnextfilename{convergence_adaptive}%
	\begin{tikzpicture}
		\pgfplotsset{table/search path={plot/data}}
		\begin{semilogyaxis}[xlabel={$\minscale;j$},x dir=reverse,
			ylabel={error},
			grid= both,
			xmin=-5.7,
			legend style={legend columns=1,at={(.98,.98)},anchor=north east},
			]
			\addplot+[konv style,draw=black,mark options={fill=black}] table[x = scalemin, y = Inf] {konvergenz_dyn_gauss_N3.dat};\addlegendentry{$\norm{\cdot}_{\infty}$}
			\addplot+[konv style,mark=triangle*,draw=black,mark options={fill=black}] table[x = scalemin, y = max] {konvergenz_dyn_gauss_N3.dat};\addlegendentry{$\maxnorm{\cdot}$}
			
			\addplot+[konv style,draw=TUDa-2b,mark options={fill=TUDa-2b,solid},densely dashed] table[x = scales, y = Inf] {konvergenz_gauss_N3.dat};\addlegendentry{$\norm{\cdot}_{\infty}$}
			\addplot+[konv style,mark=triangle*,draw=TUDa-2b,mark options={fill=TUDa-2b,solid},densely dashed] table[x = scales, y = max] {konvergenz_gauss_N3.dat};\addlegendentry{$\maxnorm{\cdot}$}
		\end{semilogyaxis}
	\end{tikzpicture}
	\fi%
	\caption{Error between analytical and numerical solution of the \ac{arm} (black, solid) and \ac{ssm} (blue, dashed) in different norms for $\maxscale=0$.}
	\label{fig:konv_adapt_gauss_N3}
\end{figure}

\begin{figure}
	\centering
	\ifpdf%
	\includegraphics{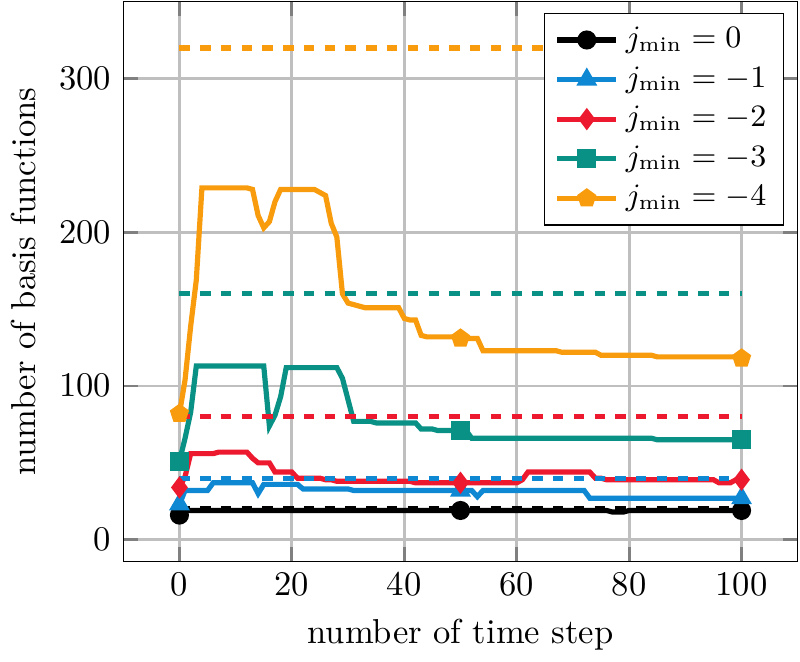}%
	\else%
	\tikzsetnextfilename{number_functions}%
	\begin{tikzpicture}
		\pgfplotsset{table/search path={plot/data}}
		\begin{axis}[xlabel={number of time step},
			ylabel={number of basis functions},
			grid= major,
			cycle list name=myCycleList,
			legend style={legend columns=1,at={(.98,.98)},anchor=north east,legend cell align=left},
			]
			\pgfplotsinvokeforeach{0,-1,...,-4}{
				\addplot+[mark repeat = 50] table[x = iteration, y = scalemin#1] {num_fct_gauss_N3.dat};\addlegendentry{$\minscale=\num{#1}$}
			};
			\pgfplotsinvokeforeach{20,40,80,160,320}{
				\addplot+[dashed,no marks] coordinates {(0,#1) (100,#1)};
			};
		\end{axis}
	\end{tikzpicture}
	\fi%
	\caption{Number of used basis functions for the \ac{arm} in every time step. The dashed lines indicate the maximum possible number of basis functions with the corresponding value of $\minscale$.}
	\label{fig:num_fct_gauss_N3}
\end{figure}

In \fig~\ref{fig:konv_adapt_gauss_N3}, the solution of the \ac{arm} is compared to those of the \ac{ssm}. The error of the former method at a minimum scale $\minscale$ has to be compared to the error of the latter at scale $j=\minscale-1$. Both methods yield the same error, as expected. However, the \ac{arm} uses less basis functions than the \ac{ssm}. \Fig~\ref{fig:num_fct_gauss_N3} shows the number of used basis functions and the maximum possible number of basis functions for every time step and for the different minimum scale. As one can see, the actual number is below the maximum number of $\L2^{-\minscale+1}$. 
\subsection{\Acl*{aq3d} method}
\begin{figure}%
	\centering
	\ifpdf%
	\includegraphics{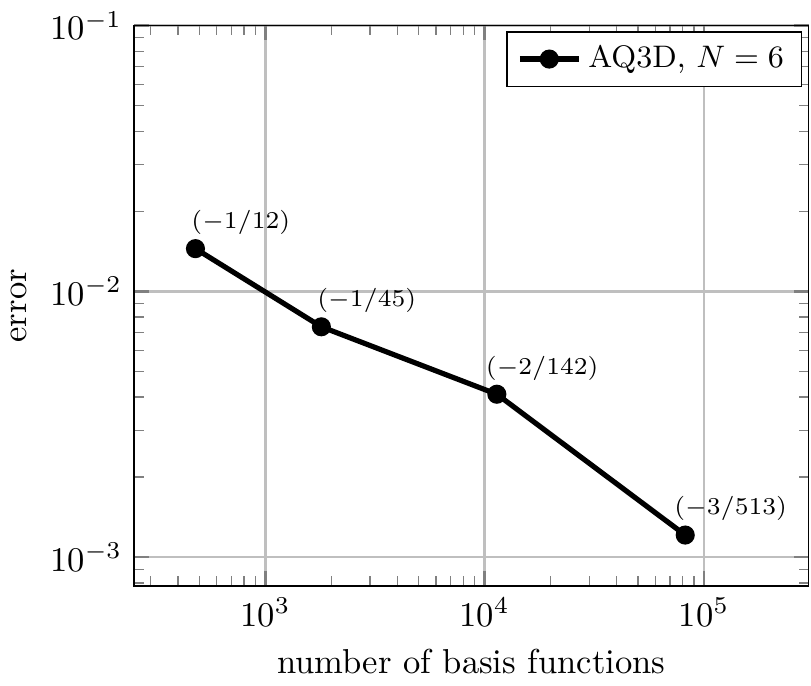}%
	\else%
	\tikzsetnextfilename{comparison_aq3d}%
	\begin{tikzpicture}
		\pgfplotsset{table/search path={plot/data/comparison}}
		\begin{loglogaxis}[konv axis,grid= major,
			xmax=3e5,
			ymax=0.1,
			xlabel={number of basis functions},
			ylabel={error},
			legend style={legend columns=1,at={(.99,.99)},anchor=north east},
			cycle list name=myCycleList,]
			\addplot+[nodes near coords={\scriptsize$(\pgfmathprintnumber[fixed,precision=0] \scale/\pgfmathprintnumber[fixed,precision=0]\anzahl)$},visualization depends on={\thisrow{scale} \as \scale},
			visualization depends on={\thisrow{anz_fe}\as \anzahl},nodes near coords style={anchor=210},] table[x=anzahl_max,y=fehler] {daten_aq3d_N6_L10_3.dat};\addlegendentry{\acs{aq3d}, $N=6$}
		\end{loglogaxis}
	\end{tikzpicture}
	\fi%
	\caption{Error between analytical and numerical solution of the \ac{aq3d} method in $\norm{\cdot}_{\infty}$.	The numbers between the parentheses are the minimum scale and the number of FE nodes.}
	\label{fig:convergence_aq3d}
\end{figure}

For verification, the \ac{aq3d} method is used to solve the problem in \eqref{eq:heat_eq_3d} with dimensionless $L=10$, $\lambda=10$, $\cv=5$, $\Delta t = \num{1e-4}$ and 10 time steps. The initial condition is taken from \eqref{eq:initial_condition} and the source term is $q(\x,t)=0$. Homogeneous Dirichlet boundary conditions are applied on the front and back side, while homogeneous Neumann boundary conditions are applied to the hull. \Fig~\ref{fig:convergence_aq3d} shows the maximum error between numerical and analytical solution over all edges in $z$-direction. The error on each edge is calculated with the infinity norm. The number of basis functions is increased by using both finer \ac{2D} meshes in the cross-section and a smaller minimum scale $\minscale$ in longitudinal direction. Furthermore, the maximum scale is set to $\maxscale=-1$ and for the spectral method, scaling functions and wavelets with $N=6$ are utilized.

Lastly, to show the potential reduction of basis functions with the \ac{aq3d} method, it is used to solve the benchmark problem with three Rutherford cables defined in \cite{DAngelo.2020}. Instead of the cables being extended \SI{1}{\meter} in $z$-direction, they are extended to \SI{10}{\meter} for this simulation. Because the center of the source term is still $\zq=\SI{0.33}{\meter}$, there are no major differences expected in the temperature distribution between $z=\SI{0}{\meter}$ and $z=\SI{1}{\meter}$. The remaining parameters are the same as in \sect~\ref{sec:results_q3d}. Moreover, scaling functions and wavelets with $N=6$, $\minscale=-4$ and $\maxscale=-3$ have been chosen. Thus, the results are comparable with the results from \sect~\ref{sec:results_q3d}. The tolerance is $\tolu=\num{1e-4}$. With the indicated values, the maximum number of basis functions in longitudinal direction is \num{320}. Furthermore, the used mesh in the cross section has \num{1289} nodes and \num{2496} elements. This leads to a maximum number of \num{412480} basis functions for the \ac{aq3d} method.

\begin{figure}
	\centering
	\ifpdf%
	\includegraphics{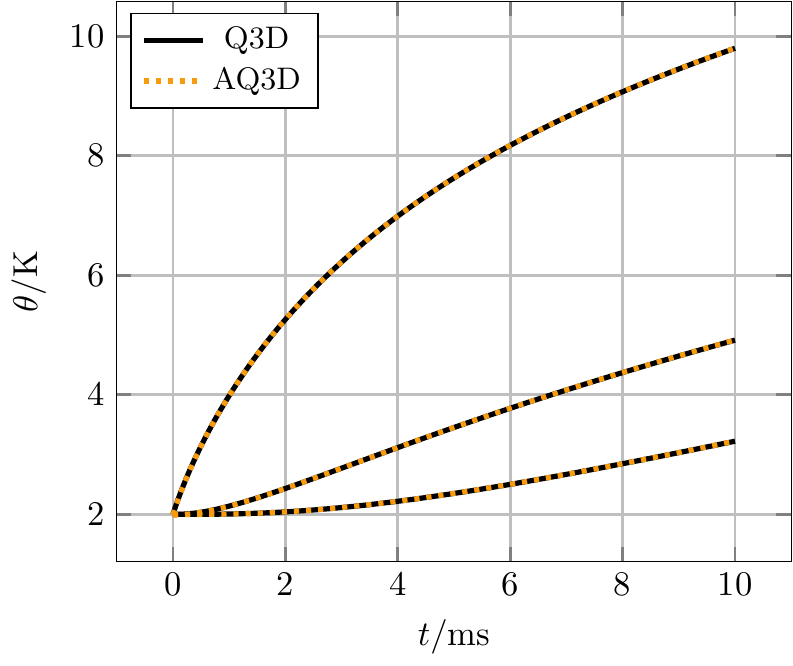}%
	\else%
	\tikzsetnextfilename{temperature_time_aq3d}%
	\begin{tikzpicture}
		\pgfplotsset{table/search path={plot/data/rutherford}}
		\begin{axis}[xlabel={$t/\si{\milli\second}$},
			ylabel={$\theta/\si{\kelvin}$},
			cycle list name=myColorCycleList,
			legend style={legend columns=1,at={(.02,.98)},anchor=north west},
			scaled x ticks = base 10:3,
			xtick scale label code/.code={\empty},
			grid=major,
			]
			\addplot+[normal plot,black] table[x=time,y=left] {rutherford_N6_temp_knoten.dat};\addlegendentry{\acs{q3d}}
			\addplot+[normal plot,black,forget plot] table[x=time,y=middle] {rutherford_N6_temp_knoten.dat};
			\addplot+[normal plot,black,forget plot] table[x=time,y=right] {rutherford_N6_temp_knoten.dat};
			\addplot+[normal plot,dotted,TUDa-7b] table[x=time,y=left] {rutherford_N6_temp_knoten_aq3d.dat};\addlegendentry{\acs{aq3d}}
			\addplot+[normal plot,dotted,TUDa-7b,forget plot] table[x=time,y=middle] {rutherford_N6_temp_knoten_aq3d.dat};
			\addplot+[normal plot,dotted,TUDa-7b,forget plot] table[x=time,y=right] {rutherford_N6_temp_knoten_aq3d.dat};
		\end{axis}
	\end{tikzpicture}
	\fi%
	\caption{Evolution of the temperature in the three cables over time calculated with the \ac{q3d} and \ac{aq3d} method. Evaluated at $z=\zq$ and in the center of the cross-section of each cable.}
	\label{fig:rutherford_temp_t_aq3d}
\end{figure}

\begin{figure}%
	\centering
	\ifpdf%
	\includegraphics{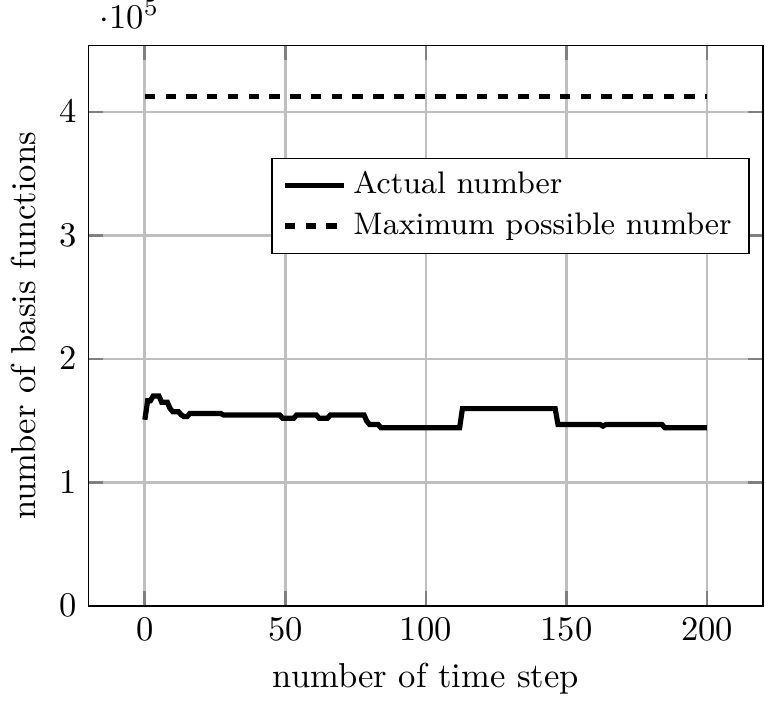}%
	\else%
	\tikzsetnextfilename{number_functions_aq3d}%
	\remake
	\begin{tikzpicture}
		\pgfplotsset{table/search path={plot/data/rutherford}}
		\begin{axis}[xlabel={number of time step},
			ylabel={number of basis functions},
			grid= major,
			ymin=0,
			cycle list name=myCycleList,
			legend style={legend columns=1,at={(.98,.8)},anchor=north east,legend cell align=left},
			]
			\addplot+[no marks] table[x = iteration, y = anzahl] {anzahl_gesamt_aq3d.dat};\addlegendentry{Actual number}
			\addplot+[black,dashed,no marks] coordinates {(0,412480) (200,412480)};\addlegendentry{Maximum possible number}
		\end{axis}
	\end{tikzpicture}
	\fi%
	\caption{Number of used basis functions for the \ac{aq3d} in every time step. The dashed line indicates the maximum possible number of basis functions with the corresponding value of $\minscale$.}
	\label{fig:num_fct_gauss_N6_aq3d}
\end{figure}

In \fig~\ref{fig:rutherford_temp_t_aq3d}, the evolution of the temperature in the three cable is shown for the \ac{q3d} and the \ac{aq3d} method. Both methods lead to the same evolution. However, the number of basis functions of the \ac{aq3d} method is significantly lower than the maximum possible number of basis functions of \num{412480} as depicted in \fig~\ref{fig:num_fct_gauss_N6_aq3d}. A simulation of this problem with the \ac{q3d} method with the same accuracy in the $z$-direction and the same mesh in the cross-section would require this maximum number of \num{412480} basis functions. 


\FloatBarrier
\section{Conclusion}\label{sec:conclusion}
A spectral method using Daubechies scaling functions has been formulated and verified. This has been combined with a \ac{2D} \ac{fem} with a nodal basis into a \acl{q3d} method. With this method, a stack of three simplified Rutherford cables has been simulated successfully. A third method with an adaptive resolution in space and time has been formulated for the \ac{1D} heat equation. This method has been extended to a \acl{q3d} method in the same manner as the spectral method. 
A comparison between the adaptive and associated non-adaptive methods showed a significant reduction of the number of basis functions, preserving the accuracy.


\begin{backmatter}

\section*{Acknowledgements}
This work has been supported by the BMBF project "Quenchsimulation für Supraleitende Magnete: Steigerung der Auflösung in Zeit und Raum" (BMBF-05P18RDRB1) and by the Graduate School Computational Engineering at TU Darmstadt. 
The research of Jonas Bundschuh is sponsored by the German Science Foundation (DFG project 436819664).

\section*{Funding}
Not applicable

\section*{Abbreviations}
\printacronyms[heading=none]

\section*{Availability of data and materials}
Not applicable

\section*{Competing interests}
The authors declare that they have no competing interests.

\section*{Authors' contributions}
The research and the numerical tests have been carried out by JB. JB and LAMD worked together on the manuscript. HDG acquired the funding and administrated the project. All authors have read and approved the final manuscript.

\printaddresses



\bibliographystyle{bmc-mathphys} 
\bibliography{literature.bib}      

\end{backmatter}
\end{document}